\documentclass[numberedappendix,apj]{emulateapj}
\usepackage{apjfonts}

\shorttitle{XMM-Newton EPIC background modeling}
\shortauthors{J. Nevalainen, M. Markevitch \& D. Lumb}
\slugcomment{Accepted for publication in ApJ}

\begin{document}

\title{XMM-Newton EPIC background modeling for extended sources}

\author{J. Nevalainen$^{1,2}$, M. Markevitch$^{1,4}$, D. Lumb$^{3}$}
\affil{Harvard-Smithsonian Center for Astrophysics, Cambridge, USA$^{1}$\\
Observatory, University of Helsinki, Finland$^{2}$\\
ESTEC, Noordwijk, Netherlands$^{3}$\\
IKI, Moscow, Russia$^{4}$}

\begin{abstract}

We use XMM-Newton blank-sky and closed-cover background data to explore the
background subtraction methods for large extended sources filling the EPIC
field of view, such as nearby galaxy clusters, for which local background
estimation is difficult.  In particular, we investigate the uncertainties of
the background modeling in the 0.8--7.0 keV band that affect
the cluster analyses.
To model the background, we have constructed composite datasets from the
blank-sky observations and compared them to the
individual blank-sky observations to evaluate the modeling error.
Our results apply to data obtained with thin and medium
optical filters and in Full frame and Extended full frame modes.

As expected, the modeling uncertainty is determined by how the EPIC
background flares are filtered.  We find that to keep this uncertainty
tolerable, one has to use a much more restrictive filter than that
commonly applied.  In particular, because flares have highly variable
spectra, not all of them are identified by filtering the $E>10$ keV light curve.
We tried using the outer part
of the EPIC FOV for monitoring the background in a softer band (1--5 keV).
We find that one needs to discard the time periods when either the
hard-band or the soft-band rate exceeds the nominal value by more than 20\%
in order to limit the 90\% CL background uncertainty to
between
$\pm5$\% at $E=4-7$ keV and $\pm20$\% at $E=0.8-1$ keV, for both MOS and
PN. This compares to a $10-30$\% respective PN uncertainty when only the
hard-band light curve is used for filtering, and to a $15-45$\% PN uncertainty 
when applying the commonly used $2-3\sigma$ filtering method.
Adding such a soft-band filter on average results in only a 5--10\%
reduction of the useful exposure time.

We illustrate our method on a nearby cluster A1795. The above background
uncertainties convert into the systematic temperature uncertainties between
$\sim\pm$1\% at $r=3-4'$ and $\pm$20--25\% ($\sim \pm$1 keV for A1795)
at $r=10-15'$.  For comparison, the commonly applied $2-3\sigma$ clipping of
the hard-band light curve misses a significant amount of flares,
rendering the temperatures beyond $r=10'$ unconstrained.
Thus, the background uncertainties do not prohibit the EPIC temperature profile 
analysis of low-brightness regions, like outer regions of galaxy clusters,
provided a conservative flare filtering such as the double filtering method with 
$\pm$20\% limits is used.

\end{abstract}

\keywords{galaxies: clusters: general --- galaxies: clusters: individual (Abell 1795) --- X-rays: galaxies}

\section{Introduction}

The total background emission in XMM-Newton EPIC instruments is known to
have a time-variable (i.e., flaring) component, in addition to the sky
background (Cosmic X-ray background + the Galactic emission) and a 
relatively quiescent cosmic ray induced internal
background components.  The first-order remedy is to filter the data using a
high energy light curve, where the particle background dominates the
emission (e.g., Arnaud et al.\ 2001; Lumb et al.\ 2002).  The hard band
filtering may miss flares that are more variable at lower energies.
When the outer parts of the FOV can be used for direct local background
estimation, the effects of any remaining flares can be minimized
(e.g.  Zhang et al., 2004).
However, the background estimation is more complicated for such extended
sources, like nearby bright clusters of galaxies, which fill the whole FOV
and thus do not allow unambiguous local background estimate to be
obtained. In this case one has to use a blank sky based background
estimate (see Lumb et al.\ 2002; Read \& Ponman 2003).
As of January 2005, XMM-Newton has observed or is scheduled to observe 
$\sim$80 nearby ($z < 0.1$) clusters of galaxies, for which the blank sky 
subtraction has to be used.

The uncertainties involved in the blank sky based background estimation may
have a significant effect on the temperature measurements of low-brightness
sources like the outer parts of clusters of galaxies.  However, these
effects have not been systematically evaluated in the works on the
XMM-Newton background published to date (e.g., Lumb et al.\ 2002; Marty et
al.\ 2003, Read \& Ponman 2003).  Consequently, the uncertainties have not
been properly propagated to most XMM-Newton cluster temperature measurements
that used blank sky backgrounds.

Our goals in this paper are to evaluate these uncertainties by comparing a
quiescent background spectrum prediction with a sample of XMM-Newton EPIC
blank sky data, and to explore ways to minimize these uncertainties. We
compare different methods of flare filtering and normalizing the blank
sky background spectrum for a given observation.  We expand the
background analysis work of Katayama et al.\ (2004), e.g. by studying the
MOS data and including more PN data and mild PN flares into the analysis.

We describe a recipe for propagating the uncertainties of the quiescent
background predictions to the cluster temperature measurements. We
illustrate this recipe by performing spatially resolved
spectroscopy of a bright nearby cluster A1795, whose XMM observations
are strongly affected by flares.

\newpage

\section{Analysis}
\label{anal}
We processed the raw data with the SAS 5.4.1 tools epchain and emchain, with the latest calibration constituents available in Jan 2004.
We selected events with patterns 0-4 (single + double) and 0-12 for PN and MOS, respectively. We used evselect tool to extract spectra, images and light curves, and the 
arfgen tool to generate auxiliary response files.
In the spectral fits we used the latest ready-made on-axis energy redistribution files. We further filtered the data with SAS expression ``flag==0'' to exclude bad
pixels and CCD gaps, and excluded the regions of bright point sources.  When
analyzing the hard band light curves or images, we select events above
energies 10 keV (PN) or 9.5 keV (MOS), up to the instrument's internal upper
limit, but when examining the hard band spectra, we limit the analysis to
energy bands 10--14 keV (PN) or 9.5--12 keV (MOS).

At the time of this
writing, there is no functional standard tool for performing the exposure
correction for the light curve in the SAS distribution, i.e. to 
correct the light curves for losses in flux due to a number of causes, e.g. 
`true' dead time, during which events are not recorded at all (see the SAS Packages 
description in the XMM-Newton web page http://xmm.vilspa.esa.es).
Thus, in order to obtain generally applicable results, we ignore this correction. Note that our reported count rates and exposure times based on the light curves
differ by $\sim$ 10\% from those derived using the spectra, which incorporate exposure correction.

We examined the PN and MOS hard band images for anomalously bright CCDs,
without finding any.  The exposure times for individual CCDs in a given
pointing in closed-cover and blank-sky samples vary by less than 1\%.  The
useful detector area of the full FOV (a circle with a radius of 15$'$,
centered at (DETX,DETY) = (-2120,-1080) for PN, (0,0) for MOS), after
exclusion of CCD gaps, bad pixels, inactive detector regions and point
sources varies from observation to observation by only 0.1\% for
each instrument, with the average values of 615 (PN) and 656 (MOS)
arcmin$^{-2}$.  These values can be used to scale our full FOV count rates,
if one excludes a significant fraction of the FOV from the analysis. 

\section{Closed-cover background}
\label{closed}

\begin{deluxetable}{llllllll}
\tabletypesize{\scriptsize}
\tablecaption{Closed-cover sample}
\tablewidth{0pt}
\tablehead{
\colhead{obs. ID} & \multicolumn{3}{c}{exp. ID}    & \colhead{obs. start} & \multicolumn{3}{c}{exp. time (ks)} \\
                  & \multicolumn{3}{c}{\hrulefill} &                      & \multicolumn{3}{c}{\hrulefill}    \\
                  & PN & MOS1 & MOS2               & year-mm-dd           & PN   & MOS1  & MOS2                }
\startdata
\multicolumn{8}{c}{FF}                                       \\   
0086360901  & -    & S005 & S006 & 2001-03-11 & -  & 10  & 10 \\
0094170301  & -    & S001 & S002 & 2001-04-12 & -  & 8   & 8  \\
0094800301  & -    & S001 & S002 & 2001-06-21 & -  & 10  & 10  \\
0108860601  & S005 & S003 & S004 & 2001-10-14 & 6  & 9   & 9   \\
0109490701  & S005 & S003 & S004 & 2001-09-07 & 4  & 7   & 7   \\
0111971501  & S005 & S003 & S004 & 2002-06-01 & 5  & 7   & 7   \\
0112830701  & S005 & S003 & S004 & 2001-12-01 & 6  & 9   & 9   \\
0122310101  & -    & -    & S002 & 2000-03-27 & -  & -   & 10  \\
0122310201  & -    & -    & S002 & 2000-03-27 & -  & -   & 34  \\
0122310401  & -    & S008 & -    & 2000-03-28 & -  & 6   & -    \\
0123720201  & -    & -    & S004 & 2000-05-01 & -  & -   & 26   \\
0123920101  & -    & -    & S014 & 2000-05-18 & -  & -   & 16   \\
0124300101  & -    & S001 & -    & 2000-05-06 & -  & 9   & -    \\
0125110101  & -    & S021 & -    & 2000-05-24 & -  & 14  & -    \\
0125320901  & -    & S015 & S016 & 2002-12-17 & -  & 10  & 10   \\
0125910201  & -    & S007 & S008 & 2002-06-17 & -  & 13  & 13   \\
0129321301  & S002 & -    & -    & 2000-08-22 & 7  & -   & -    \\
0134521601  & S005 & -    & -    & 2002-06-18 & 21 & -   & -    \\
0136540501  & S008 & S003 & S006 & 2002-11-04 & 19 & 22  & 22   \\
0136750301  & U002 & U002 & U002 & 2001-05-22 & 25 & 30  & 30   \\
0154150101  & S003 & U002 & U003 & 2002-01-31 & 18 & 24  & 22   \\
            &      &      &      &            &    &     &      \\
\multicolumn{8}{c}{EF}                    \\
0134521701  & S005 & -    & -    & 2002-11-15 & 15 &  -  & -    \\
0134720401  & S005 & -    & -    & 2002-10-05 & 10 &  -  & -    \\
0160362601  & S005 & -    & -    & 2003-08-02 & 18 &  -  & -    \\
\enddata
\tablecomments{Exp. time shows LIVETIME values for the central CCD and are shown only for the sets included in the sample. 
FF and EF show the Full frame and Extended full frame subsets}
\label{t1.tab}
\end{deluxetable}

In order to estimate the spectrum of the quiescent particle-induced
background, we formed a sample of EPIC data obtained with the filter wheel
in closed position (i.e., when no photons from the mirror enter the
detectors), see Table \ref{t1.tab}. The sample consists of all the
publicly available closed-filter Full frame or Extended full frame data sets
as of Oct 2003, with livetime for the central CCD larger than 5ks in any of
the detectors, as reported in the observation log browser in the XMM-Newton
web page. We excluded sets which have reported problems
of high radiation.  In some observations there are periods of high radiation
in the beginning and end of the exposure, and we excluded those periods when
extracting spectra.  Using the hard band ($>$ 10 keV for PN and $>$ 9.5 keV
for MOS) light curves of the full FOV closed-cover data in each individual
observation, we obtain count rates which are scattered around the sample
averages of 0.63 (PN), 0.17 (MOS1) and 0.18 (MOS2) cts s$^{-1}$ with a
standard deviation of 6\%.

\begin{figure}
\plotone{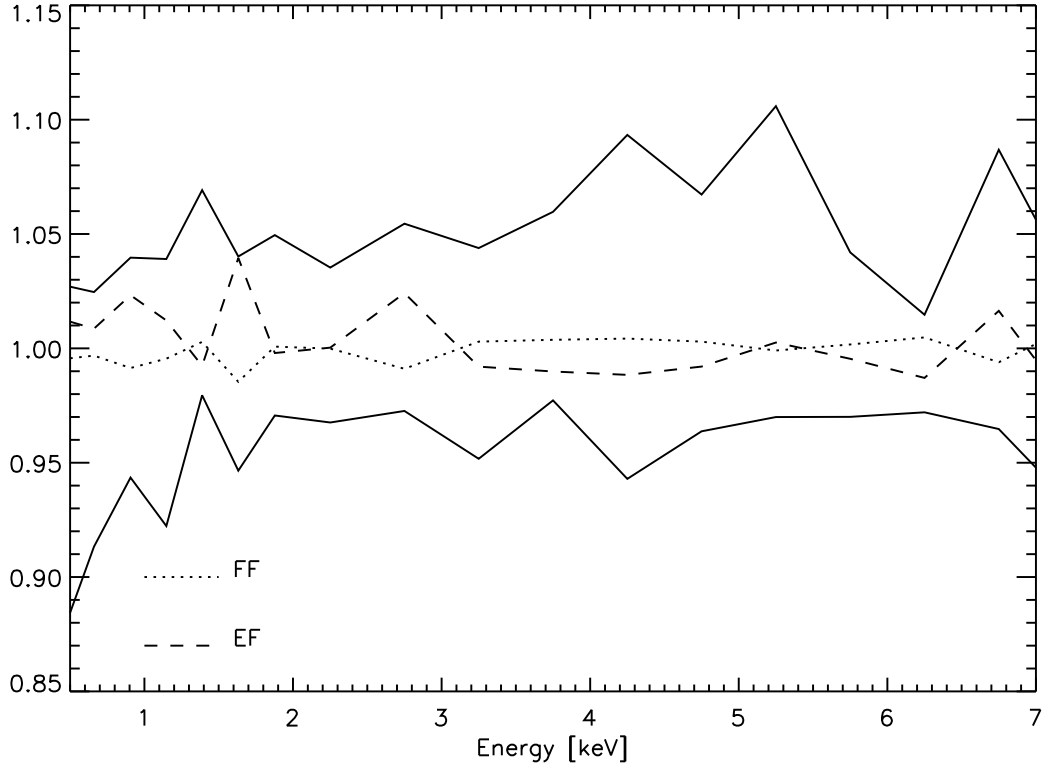}
\caption[]{Solid lines show the relative deviation of the individual
  renormalized closed-cover PN spectra from the average at 90\% CL.  The
  ratio of the renormalized average Full frame (FF) and Extended full frame
  (EF) spectra to the sample average are shown as dotted and dashed lines,
  respectively}
\label{f1.fig}
\end{figure}

We formed the average closed-cover spectra by co-adding the individual
spectra. The exposure times for the combined spectra are 110 ks,
40 ks, 170 ks and 220 ks for PN Full frame, PN Extended full frame, MOS1 and
MOS2, respectively.  In order to examine the variation of the spectral shape
among the individual pointings, we renormalized the individual spectra so
that their hard-band (10--14 keV for PN, 9.5--12 keV for MOS) count
rates are the same as in the co-added spectrum.  To avoid statistical
scatter, we binned the spectra to better than 10\% statistical accuracy in
each bin. In the interesting 0.8--7.0 keV band (used below) the
individual pointing spectra are consistent within 10\% for each
instrument (see Fig.\ \ref{f1.fig} for PN).  This stability is
encouraging considering the background modeling. Furthermore, the
co-added, renormalized spectra for the PN Full frame and Extended full
frame modes are consistent with each other within a few \% in the 0.8--7.0
keV band (Fig.  \ref{f1.fig}).  Thus, in the following we combine
them in the single PN closed-cover spectrum.

The co-added PN (MOS) closed cover spectrum can be adequately fitted in the
0.5--14.0 (0.5--12.0) keV band with a model consisting of a broken power-law
continuum and several Gaussians for the instrumental lines (see Fig.
\ref{f2.fig}). For this fit, the model was not multiplied by the
instrument effective area, only convolved with the energy
redistribution matrix.  At the low energies the photon indices and break
energies of PN and MOS are similar ($\alpha_{soft} \sim 0.7-0.8$ ,
E$_{break} \sim 1.3-1.5$ keV), while MOS spectra harden slightly more
towards higher energies ($\alpha_{hard} = 0.4$ for PN and 0.1--0.2 for MOS).

\section{Flare filtering}

The total EPIC background includes strong time-variable particle-induced
component, i.e. flares.  In the following we explore the flare filtering
methods using the blank-sky data.

\label{quies}

\begin{deluxetable*}{llllllllllll}
\tabletypesize{\scriptsize}
\tablecaption{Blank sky sample}
\tablewidth{0pt}
\tablehead{
\colhead{obs. ID} & \colhead{obs. start} & \colhead{t$_{exp}$} & \colhead{RA} & \colhead{DEC} & \multicolumn{3}{c}{filter} &  \multicolumn{2}{c}{mode} & \colhead{NH}
 & \colhead{field}\\
\colhead{} &  \colhead{} & \colhead{} & \colhead{} & \colhead{}  & \multicolumn{3}{l}{\hrulefill} & \multicolumn{2}{l}{\hrulefill}  & \colhead{}  & \colhead{}  \\
\colhead{}                 & \colhead{year-mm-dd}  &  \colhead{ks}   & \colhead{(J2000)} & \colhead{(J2000)}  & \colhead{PN} & \colhead{M1} & \colhead{M2} & \colhead{PN} & \colhead{MOS} & \colhead{$10^{20}$ cm$^{-2}$} & \colhead{} }
\startdata
0098810101 & 2000-12-04 & 19  & 39.048  & -52.321 & T & T & T  & FF & FF & 3.0 & WW Hor              \\
0106660201 & 2000-11-18 & 45  & 333.882 & -17.735 & T & T & T  & FF & FF & 2.4 & LBQS 2212-1759      \\
0106660601 & 2001-11-17 & 96  & 333.882 & -17.735 & T & T & T  & FF & FF & 2.4 & LBQS 2212-1759      \\
0108060401 & 2001-07-27 & 32  & 53.117  & -27.808 & T & T & T  & EF & FF & 0.9 & AXAF Ultra Deep F   \\
0108060501 & 2001-07-27 & 43  & 53.117  & -27.808 & T & T & T  & EF & FF & 0.9 & AXAF Ultra Deep F   \\
0108060601 & 2002-01-13 & 53  & 53.117  & -27.808 & T & T & T  & EF & FF & 0.9 & AXAF Ultra Deep F   \\
0108060701 & 2002-01-14 & 78  & 53.117  & -27.808 & T & T & T  & EF & FF & 0.9 & AXAF Ultra Deep F   \\
0108061801 & 2002-01-16 & 51  & 53.117  & -27.808 & T & T & T  & EF & FF & 0.9 & AXAF Ultra Deep F   \\
0108061901 & 2002-01-17 & 41  & 53.117  & -27.808 & T & T & T  & EF & FF & 0.9 & AXAF Ultra Deep F   \\
0108062101 & 2002-01-20 & 50  & 53.122  & -27.806 & T & T & T  & EF & FF & 0.9 & AXAF Ultra Deep F   \\
0108062301 & 2002-01-23 & 75  & 53.117  & -27.808 & T & T & T  & EF & FF & 0.9 & AXAF Ultra Deep F   \\
0109660801 & 2001-06-12 & 57  & 203.654 & 37.912  & T & T & M  & FF & FF & 0.8 & Deep Field 1334+37  \\
0111550101 & 2001-05-18 & 38  & 189.208 & 62.217  & T & T & T  & FF & FF & 1.5 & Hubble Deep Field N \\
0111550201 & 2001-05-18 & 37  & 189.208 & 62.220  & T & T & T  & FF & FF & 1.5 & Hubble Deep Field N \\
0111550401 & 2001-06-01 & 83  & 189.237 & 62.225  & T & T & T  & FF & FF & 1.5 & Hubble Deep Field N \\
0112370101 & 2000-07-31 & 48  & 34.500  & -5.000  & T & T & T  & FF & FF & 1.5 & SDS-1               \\
0112370301 & 2000-08-04 & 54  & 34.900  & -5.000  & T & T & T  & FF & FF & 1.5 & SDS-2               \\
0123700101 & 2000-04-27 & 40  & 163.179 & 57.480  & T & T & TK & FF & FF & 0.6 & Lockman Hole        \\
\enddata
\tablecomments{Exposure times are livetimes for the central PN CCD. ``T'', ``M'' and ``TK'' denote the thin, medium, and thick optical filters. EF and FF denote the Extended full frame mode and the Full
frame mode}
\label{t2.tab}
\end{deluxetable*}

\subsection{Blank sky data}
Our blank sky data sample (Table \ref{t2.tab}) consists of 18 pointings
with $N_H \le 3 \times 10^{20} {\rm cm}^{-2}$ outside of the Galactic
Spur with exposures longer than 20 ks, in which the resolved sources
occupy an insignificant fraction ($<$ 1\%) of the FOV.

Most of the blank sky observations in the sample are obtained with the
``thin'' optical filter (Ehle et al., 2004). We will restrict our analysis to the
0.8--7.0 keV band, in which the transparency of the thin filter
differs from that of the ``medium'' filter by less than 2\%. Thus, our
results are applicable also to the medium filter data.

The PN sample consists of observations performed both in the Full frame and
the Extended full frame readout modes. As mentioned above, the total internal
background spectrum is consistent between the two modes. The total
photon background is mainly affected by the telescope (effective area),
filter (transmission) and CCD (quantum efficiency) which do not change
between the PN modes. The slightly different uncorrected (see below)
contribution of out-of-time events (Ehle et al., 2004) yields a negligible 
difference in the normalization of the total blank sky spectra 
between the Full frame and Extended full frame modes.  Thus we do not separate 
the data in different PN modes in the analysis below, and our results are 
applicable to PN data obtained in both PN modes.

\subsection{Hard band light curves}
\label{hardlc}

\begin{deluxetable*}{lllllllllllll}
\tabletypesize{\scriptsize}
\tablecaption{Quiescent levels}
\tablewidth{0pt}
\tablehead{
\colhead{}        & \multicolumn{4}{c}{PN} & \multicolumn{4}{c}{MOS1} &  \multicolumn{4}{c}{MOS2} \\
\colhead{}        & \multicolumn{4}{l}{\hrulefill} & \multicolumn{4}{l}{\hrulefill} & \multicolumn{4}{l}{\hrulefill} \\
 \colhead{}       &  \multicolumn{2}{c}{hard} & \colhead{soft} & \colhead{double} &  \multicolumn{2}{c}{hard} & \colhead{soft} & \colhead{double} & \multicolumn{2}{c}{hard} & \colhead{soft} & \colhead{double} \\
\colhead{}        & \multicolumn{2}{l}{\hrulefill} & & & \multicolumn{2}{l}{\hrulefill} & & &  \multicolumn{2}{l}{\hrulefill} & & \\
\colhead{obs. ID} & \colhead{cr} & \colhead{t$_{frac}$} & \colhead{cr} & \colhead{t$_{frac}$} & \colhead{cr} & \colhead{t$_{frac}$} 
& \colhead{cr} & \colhead{t$_{frac}$} & \colhead{cr} & \colhead{t$_{frac}$} & \colhead{cr} & \colhead{t$_{frac}$} \\
\colhead{} & \colhead{cts s$^{-1}$} & \colhead{} & \colhead{cts s$^{-1}$} & \colhead{} & 
             \colhead{cts s$^{-1}$} & \colhead{} & \colhead{cts s$^{-1}$} & \colhead{} & 
             \colhead{cts s$^{-1}$} & \colhead{} & \colhead{cts s$^{-1}$} & \colhead{} }
\startdata
0098810101 & 0.59  & 0.95 & 0.28  & 0.86 & 0.15  & 0.96 & 0.14  & 0.96 & 0.16  & 0.96 & 0.14  & 0.96 \\
0106660201 & 0.63  & 0.57 & 0.28  & 0.50 & 0.18  & 0.66 & 0.14  & 0.49 & 0.18  & 0.74 & 0.14  & 0.53 \\
0106660601 & 0.72  & 0.66 & 0.30  & 0.59 & 0.19  & 0.75 & 0.15  & 0.69 & 0.20  & 0.75 & 0.15  & 0.67 \\
0108060401 & 0.62  & 0.35 & 0.28  & 0.31 & 0.18  & 0.38 & 0.14  & 0.35 & 0.19  & 0.35 & 0.14  & 0.32 \\
0108060501 & 0.63  & 0.53 & 0.27  & 0.47 & 0.19  & 0.65 & 0.14  & 0.54 & 0.18  & 0.60 & 0.14  & 0.49 \\
0108060601 & 0.72  & 0.70 & 0.29  & 0.62 & 0.21  & 0.74 & 0.15  & 0.74 & 0.21  & 0.71 & 0.15  & 0.69 \\
0108060701 & 0.76  & 0.76 & 0.29  & 0.68 & 0.21  & 0.77 & 0.15  & 0.74 & 0.22  & 0.78 & 0.15  & 0.77 \\
0108061801 & 0.70  & 0.58 & 0.28  & 0.51 & 0.20  & 0.71 & 0.14  & 0.60 & 0.20  & 0.72 & 0.14  & 0.62 \\
0108061901 & 0.64  & 0.78 & 0.27  & 0.69 & 0.18  & 0.79 & 0.14  & 0.75 & 0.19  & 0.79 & 0.14  & 0.79 \\
0108062101 & 0.62  & 0.66 & 0.27  & 0.58 & 0.17  & 0.82 & 0.14  & 0.78 & 0.19  & 0.84 & 0.14  & 0.82 \\
0108062301 & 0.61  & 0.82 & 0.25  & 0.73 & 0.17  & 0.81 & 0.14  & 0.76 & 0.18  & 0.82 & 0.14  & 0.81 \\
0109660801 & 0.66  & 0.50 & 0.28  & 0.44 & 0.18  & 0.58 & 0.14  & 0.52 & 0.19  & 0.61 & 0.14  & 0.54 \\
0111550101 & 0.68  & 0.81 & 0.30  & 0.72 & 0.18  & 0.76 & 0.13  & 0.74 & 0.18  & 0.76 & 0.14  & 0.76 \\
0111550201 & 0.68  & 0.64 & 0.28  & 0.57 & 0.18  & 0.67 & 0.13  & 0.67 & 0.18  & 0.72 & 0.13  & 0.70 \\
0111550401 & 0.69  & 0.82 & 0.29  & 0.72 & 0.18  & 0.81 & 0.13  & 0.79 & 0.18  & 0.80 & 0.14  & 0.76 \\
0112370101 & 0.62  & 0.54 & 0.26  & 0.47 & 0.16  & 0.66 & 0.13  & 0.53 & 0.17  & 0.69 & 0.13  & 0.50 \\
0112370301 & 0.61  & 0.38 & 0.27  & 0.33 & 0.17  & 0.42 & 0.14  & 0.34 & 0.18  & 0.36 & 0.13  & 0.30 \\
0123700101 & 0.67  & 0.45 & 0.29  & 0.40 & 0.20  & 0.67 & 0.16  & 0.63 & 0.20  & 0.53 & -     & -    \\
           &       &      &       &      &       &      &       &      &       &      &       &      \\
mean       & 0.669 & 0.63 & 0.281 & 0.55 & 0.184 & 0.70 & 0.140 & 0.66 & 0.189 & 0.69 & 0.140 & 0.66  \\
\enddata
\tablecomments{``cr'' shows the mean count rates of the blank-sky sample in
the full FOV in the hard band (``hard'') and in the 12--15 arcmin annulus in
the soft band (``soft''), in the faint end (i.e., the individual quiescent
levels). The count rate ranges used to determine the quiescent levels for
the hard and soft bands are 0.54--0.80 cts s$^{-1}$ and 0.22--0.34 cts s$^{-1}$
for PN; 0.12--0.24 cts s$^{-1}$ and 0.10--0.18 cts s$^{-1}$ for MOS.  Note that
the count rate values don't include the exposure correction.  t$_{frac}$
shows the fraction of useful exposure to total time after applying the
hard-band $\pm$20\% filter, or with the double filter (``double'').}
\label{t3.tab}
\end{deluxetable*}

The standard method of selecting the quiescent periods for scientific
analysis is to monitor the highest energies where the effective area for
sky photons is very low (e.g., Arnaud et al.\ 2001; Lumb et al.\ 2002).
In many XMM-Newton works, a fixed number (2--3) of standard deviations of
the hard-band count rate distribution (e.g., Pratt \& Arnaud (2002, 2003);
Reiprich et al. 2003), or a fixed upper hard band count rate limit (e.g.
Arnaud et al.\ 2001, 2002; Gastaldello et al.\ 2003) is used to accumulate
the quiescent spectra.  In order to obtain a large enough number of time
bins for an adequate distribution, the time bin size in the above works is
small, typically 100s. The resulting number of counts per bin is of
the order of 10, and thus the statistical uncertainties of each bin are
several 10\%. Thus, the 2--3 $\sigma$ clipping corresponds to an upper
filtering limit of 1.5--2.0 times the quiescent level, allowing possible
low-flux flare contamination at this level.  

It is entirely expected, and will be shown below, that such a background
modeling uncertainty of up to a factor of 2 is too big for any accurate
analysis of the low-brightness objects. Therefore, we instead use a fixed
fraction of $\pm$20\% of the quiescent level for each pointing to define the
hard band filtered quiescent periods GTI$_h$.  The adopted limit of
$\pm$20\% is a compromise between minimizing the allowed background
variability and maximizing the final exposures. In order to minimize the
loss of quiescent counts due to statistical uncertainties exceeding the 20\%
limit, we attempted to maximize the number of counts in each time bin. To
this end, we used a time bin size of 1 ks.  A possible drawback of
using bigger time bin size is that very short flares will not be detected.
  
\begin{figure}
\epsscale{0.90}
\plotone{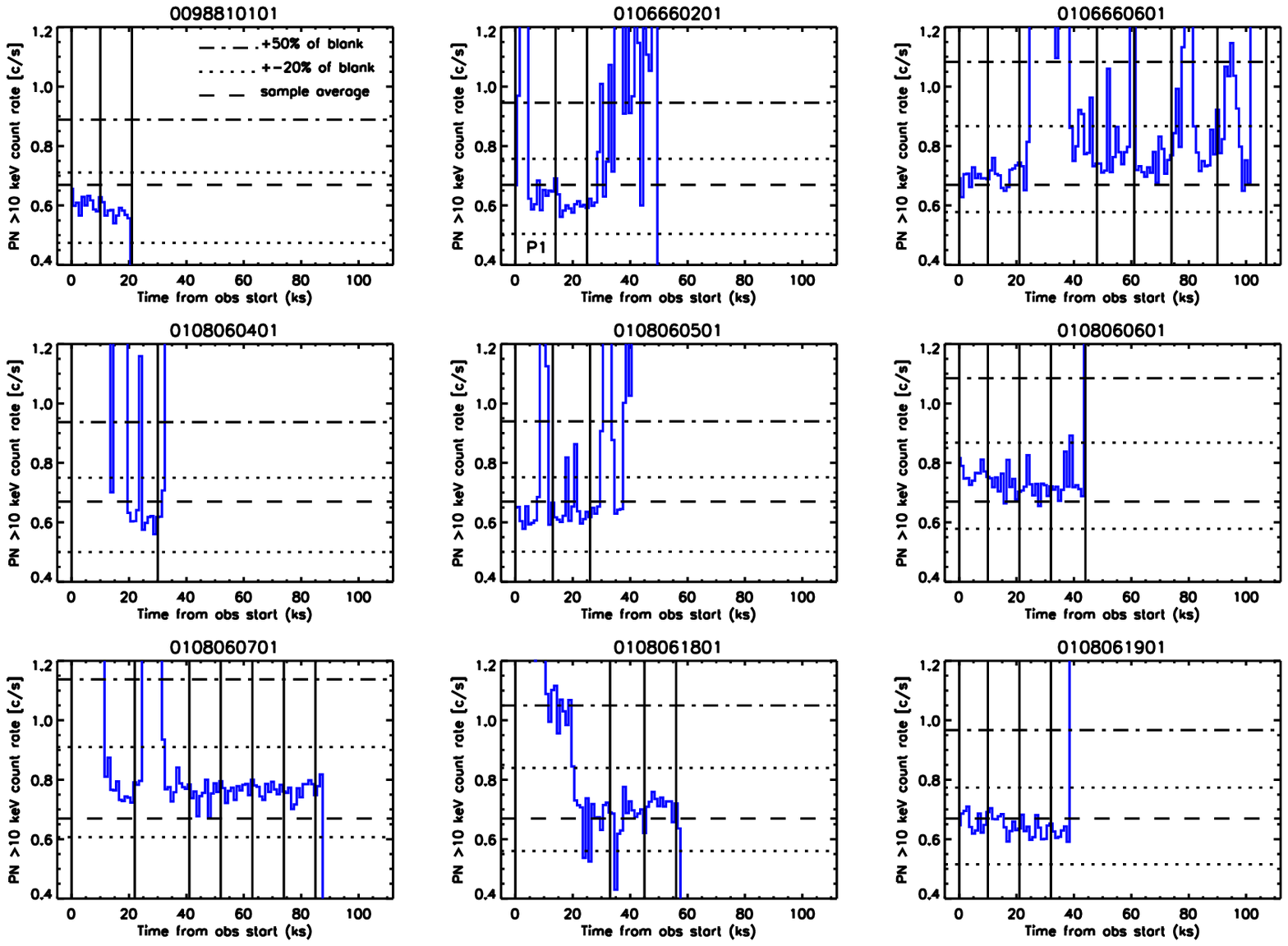}
\plotone{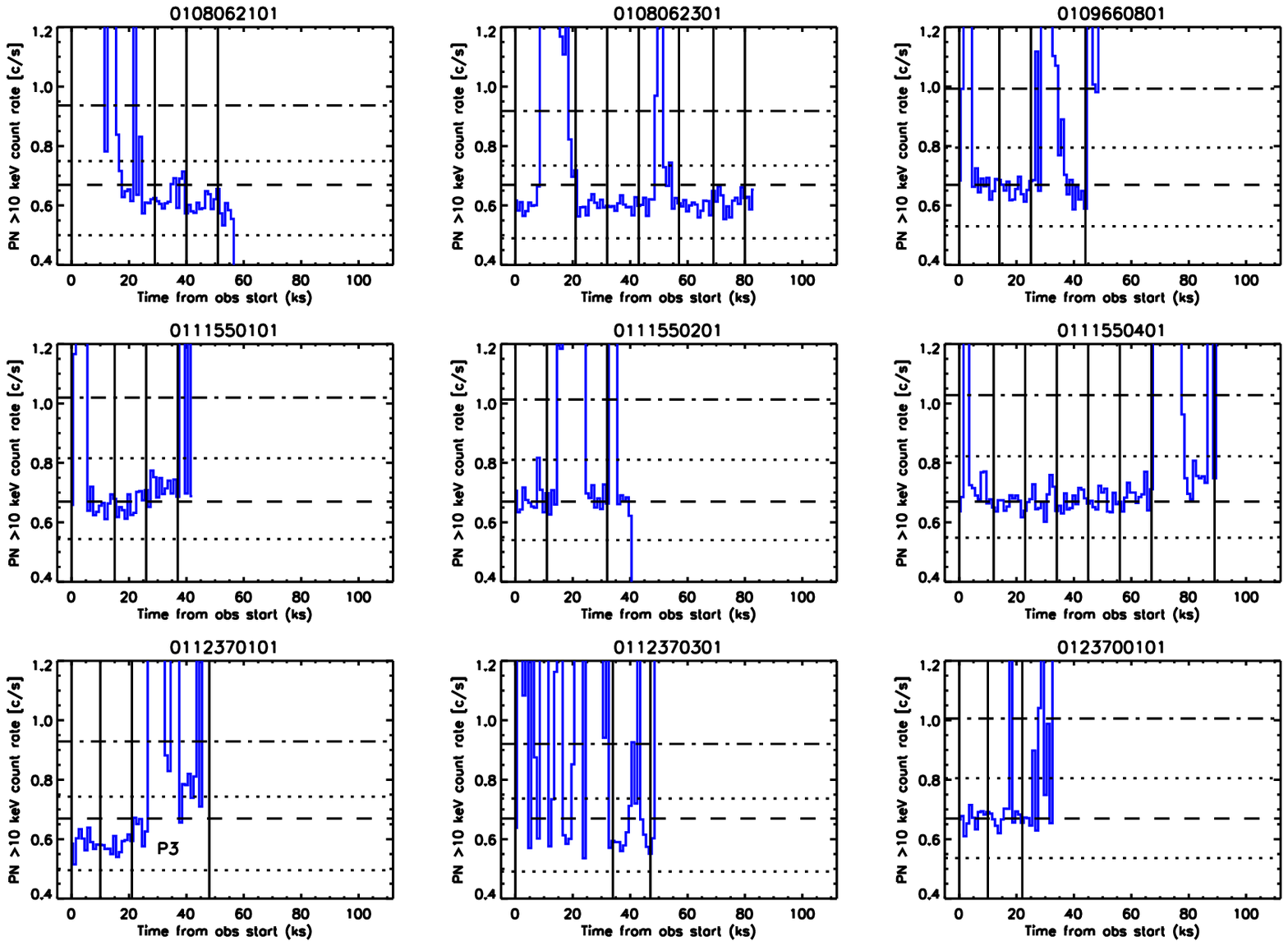}
\caption[]{EPIC PN $E>10$ keV, full-FOV light curves in 1 ks bins for the
  observations listed in Table \ref{t2.tab}, together with the
  sample-average quiescent rate (dashed line), the $\pm$ 20\% limits around
  the individual mean (dotted line), and a level exceeding the quiescent by
  50\% (dash-dot line). The vertical lines show the start and stop times of
  the 10 ks periods used in the analysis. P1 (in 0106660201) and P3 (in
  0112370101) mark the specific periods discussed in the text.}
\label{f3.fig}
\end{figure}

We also experimented with different choices for patterns and energy bins,
and noted that a more strict set of choices (i.e. only single patterns and
10--12 keV band) only serves to reduce the number of counts with no improvement in
the scatter of the renormalized blank sky spectra (see below). We thus chose
to use the channels above 10 keV (PN) or 9.5 keV (MOS), up to the
instrument's internal upper limit, when extracting the hard band light
curves (see Fig. \ref{f3.fig} for PN).  With these choices, the
number of counts per bin is a few hundreds, and thus our adopted
$\pm$20\% quiescence limits contain 2--3 $\sigma$ of the rate
distribution, so the statistical scatter above and below this threshold is
negligible.
In \S\ref{approxim} we will examine the effect of using a higher
limiting factor, similar to the commonly used filtering.

\begin{figure}
\epsscale{0.90}
\plotone{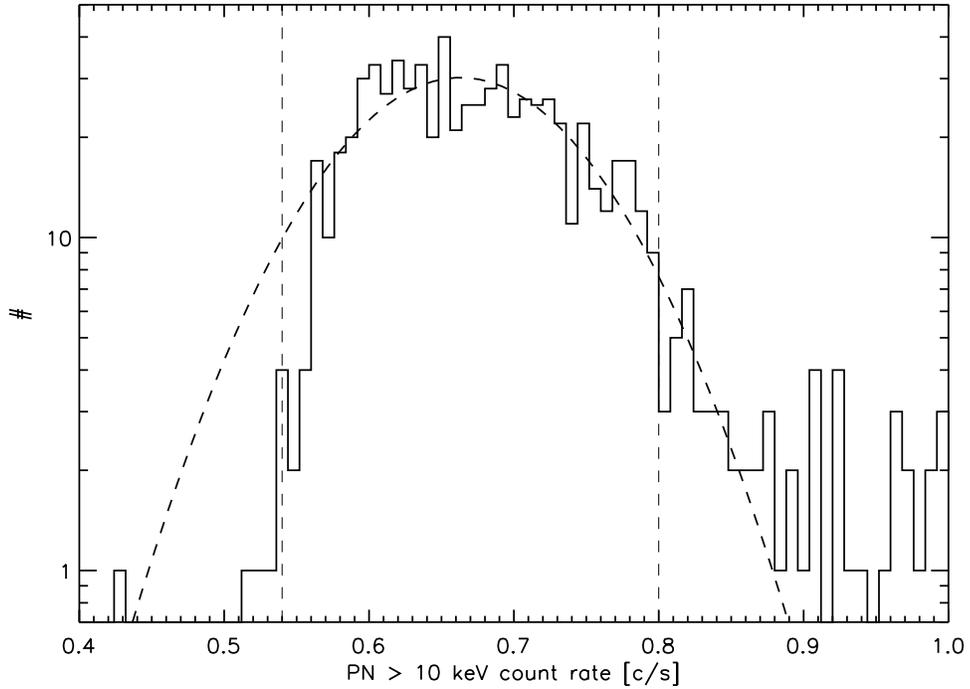}
\caption[]{The distribution of PN $E>10$ keV counts of the blank-sky sample
  in 1 ks time bins, with a best-fit Gaussian (dashed curve).  The range
  chosen to determine the quiescent levels in the blank sky pointings
  (0.54--0.80 cts s$^{-1}$) is shown by vertical dashed lines.}
\label{f4.fig}
\end{figure}

Our hard-band light curves show a wide variety of flaring activity, from
none (ID 0098810101) to 60\% of the total exposure time (ID 0112370301)
affected by flares (see Fig. \ref{f3.fig}).  The distribution
of the count rates of the whole sample peaks at the low values, (see Fig.
\ref{f4.fig}), implying a comparable quiescent level in all
the pointings.  The statistical errors of the rates in the 1ks time bins 
are 4--7\%, while the centroids of the fitted Gaussians to the rate distributions 
of individual pointings vary by 10--15\%. This implies
true variation of the quiescent flux with time at this level, presumably due
to the variation of the cosmic ray intensity and/or the particle background
(see De Luca and Molendi, 2004).
Considering this variation, we chose to define the hard-band quiescent level as the 
mean (not exposure corrected, see \S\ref{anal}) count rate in the faint end 
ranges of 0.540--0.800 cts s$^{-1}$ for PN and 0.120--0.240 cts s$^{-1}$ for MOS
for each pointing.
We used the
$\pm$20\% limits around these levels to select the quiescent periods GTI$_h$
(as explained above).  Considering the full sample as a single long
exposure, the average values in the above count rate ranges for PN, MOS1 and
MOS2 are 0.669 cts s$^{-1}$, 0.184 cts s$^{-1}$ and 0.189 cts s$^{-1}$ in the hard
band.  These values are consistent with the non-simultaneous closed cover
sample averages within the scatter (see \S\ref{closed}).

The fraction of useful hard band filtered exposure time of the total sample
is $\sim$65\% ($\sim$70\%) for PN (MOS), and it varies between 40\% --
100\% in individual pointings.  The co-added exposure times for the average
blank sky spectrum (including exposure correction) are 630ks, 780ks and
750ks for PN, MOS1 and MOS2.

\subsection{Mild flares}
\label{mildflare}

\begin{figure}
\epsscale{0.90}
\plotone{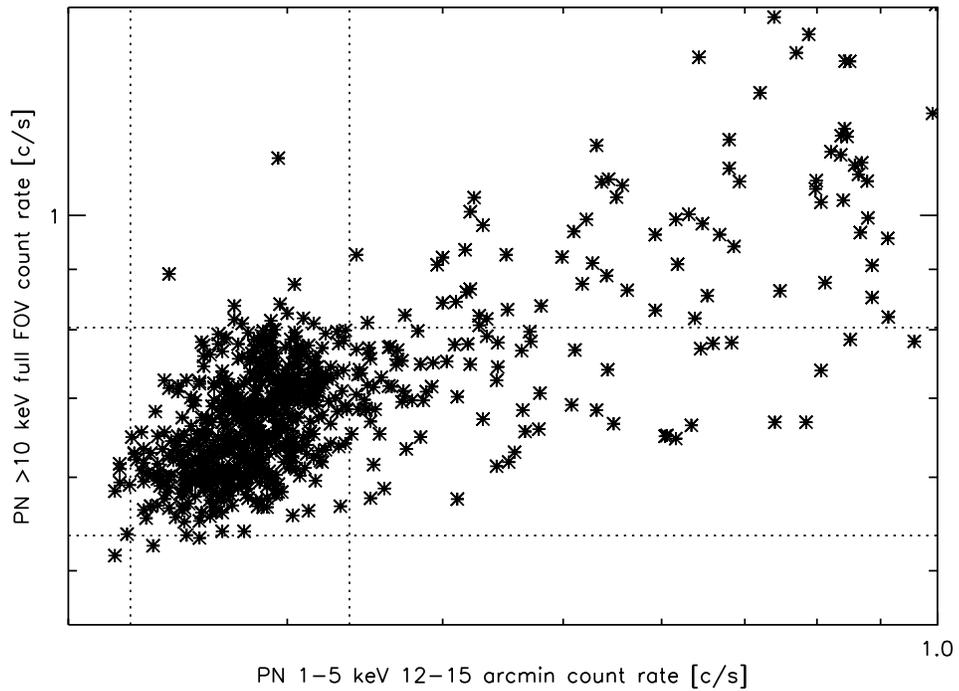}
\caption[]{The faint end of the not flare-filtered PN
  count rates of the blank-sky sample in 1 ks time bins, in the 1--5
  keV band for the 12--15 arcmin annuli and in the $E>10$ keV band for the
  full FOV.  Our adopted ranges for determining the quiescent levels in the
  blank sky pointings (0.22--0.34 cts s$^{-1}$ in the soft band; 0.54--0.80 c
  s$^{-1}$ in the hard band) are shown by dotted lines.}
\label{f5.fig}
\end{figure}

\begin{figure}
\plotone{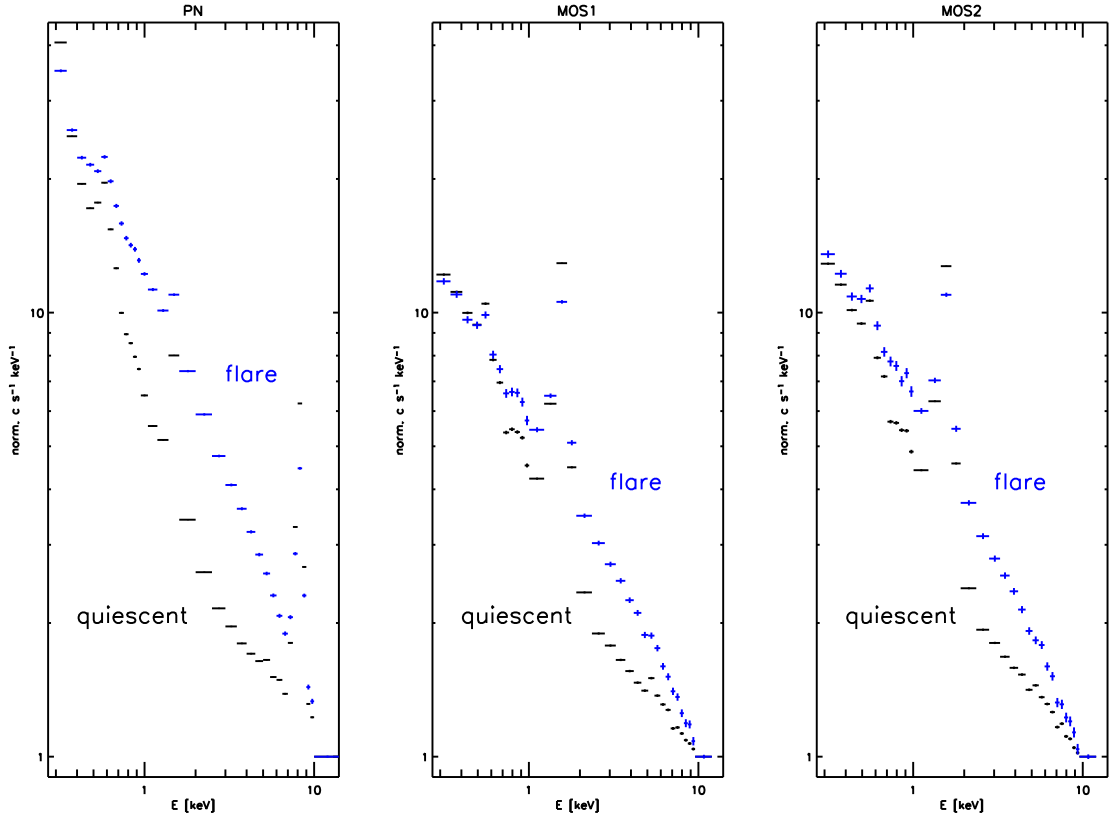}
\caption[]{The co-added, hard-band-filtered quiescent full FOV spectra
  (black) and the ``mild flare'' spectra (blue), renormalized to 1 in
  the hard band. The mild flare spectra are selected using periods when
  the hard-band count rate exceeded the quiescent level by a factor between
  1.2 and 2.}
\label{f6.fig}
\end{figure}

However, the exclusion of time intervals with elevated high-energy
background rate turns out insufficient in many observations.  During some
quiescent periods, as determined by the hard band, the PN 1--5 keV count
rate is high, reaching a factor of 3 times the quiescent level (see Fig.
\ref{f5.fig}).  To illustrate the spectra of the flares that may pass
the high-energy filter, we selected the time periods GTI$_{flare}$ with
flares that are just above our factor of 1.2 cutoff but within a factor of 2
of the quiescent rate. The resulting co-added full FOV mild flare spectra
of the whole sample, following renormalization of the hard band count rate to unity,
are substantially different from the quiescent spectra (Fig.\ 
\ref{f6.fig}). In particular, the mild-flare spectra are higher than
the quiescent spectra by different factors at different energies, up to
factors of 2.4 (PN) and 1.7 (MOS) at 2--3 keV.

\begin{figure}
\plotone{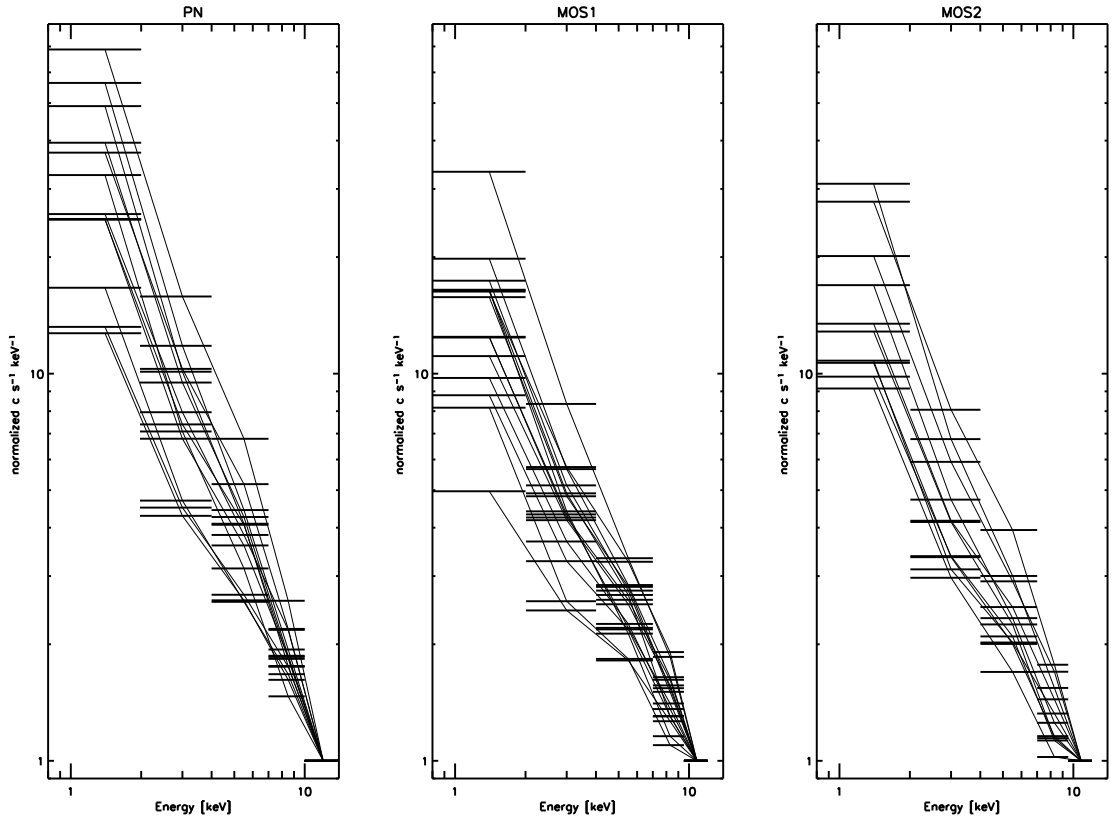}
\caption[]{The mild-flare excess spectra (i.e., the mild flare spectra minus
  the quiescent spectra, see \S\ref{mildflare}) normalized to unity at
  10--14 keV (PN, left panel) or at 9.5--12 keV (MOS1 and MOS2, middle and
  right panels). The quiescent spectra are obtained by the $\pm$20\%
  filter while the mild-flare periods are selected by applying the factor
  1.2-2 filter in the hard band (as in Fig.\ \ref{f6.fig}). The time
  intervals are selected in such a way that each spectrum has about 5 ks
  exposure.}
\label{f7.fig}
\end{figure}

To study the temporal stability of the above mild flare spectra, we
split the individual pointings into $\sim$10 periods containing $\sim$5 ks
of mild flare exposure. In order to ensure useful signal-to-noise ratio,
we used large spectral bins of 0.8--2--4--7 keV.  By subtracting the
co-added blank sky spectra from the 5 ks mild flare spectra, we
obtained the mild flare excess spectra, which we then renormalized to yield
unity count rate in the hard band.  Comparison of the renormalized mild
flare excess spectra shows (Fig. \ref{f7.fig}) that the flares have
highly variable spectral shape. The individual 5 ks spectra differ from each other
more at lower energies, by a factor of 3--7 in maximum at 0.8--2.0 keV 
in different instruments.

Note that the full range of the flares shown in Fig.\ \ref{f7.fig}
would pass the hard-band filter if the upper limit is 2 times the
quiescent level, which corresponds to that in many works that use the
2--3 $\sigma$ clipping method.  Thus, a more conservative filtering
is necessary to select the quiescent data in order to minimize the
time-variable flare component, which can give very different and
unpredictable contribution in any given observation.  The increasing
temporal variability of the spectral shape of the mild flares towards the
low energies indicates the need for better, soft-band based filtering (see
also, e.g., Pointecouteau et al.\ 2004).  We will do this in \S\ref{softfil}.  If one wants to use the data obtained during mildly
flaring periods, a more detailed method than simple rescaling of the
quiescent background would be needed, which we will not attempt in the
current work.

\subsection{Soft-band light curves}
\label{softfil}

The obvious problem in using lower energies for flare filtering when
analyzing a bright extended source, like a galaxy cluster, is the dominance
of the source emission.  However, the decline of the cluster brightness, as
well as the decrease of the effective area with radius reduces the cluster
contribution at the edge of the FOV. For example, for the bright nearby
cluster A1795 in the 1--5 keV band, the cluster contribution in the
$r=12-15'$ detector region is less than 50\% of the total emission (see
\S\ref{a1795lc}).  The absolute values of the $\pm$ 20\% variation of
the blank sky quiescent level (see below) are thus 10--15\% of the total
quiescence level (A1795 plus the background) in the 1--5 keV band, or,
greater than 2 times the statistical uncertainties. Thus, the soft-band
filtering should be applicable to most clusters, except for the nearest ones
like Coma.

\begin{figure}
\epsscale{0.90}
\plotone{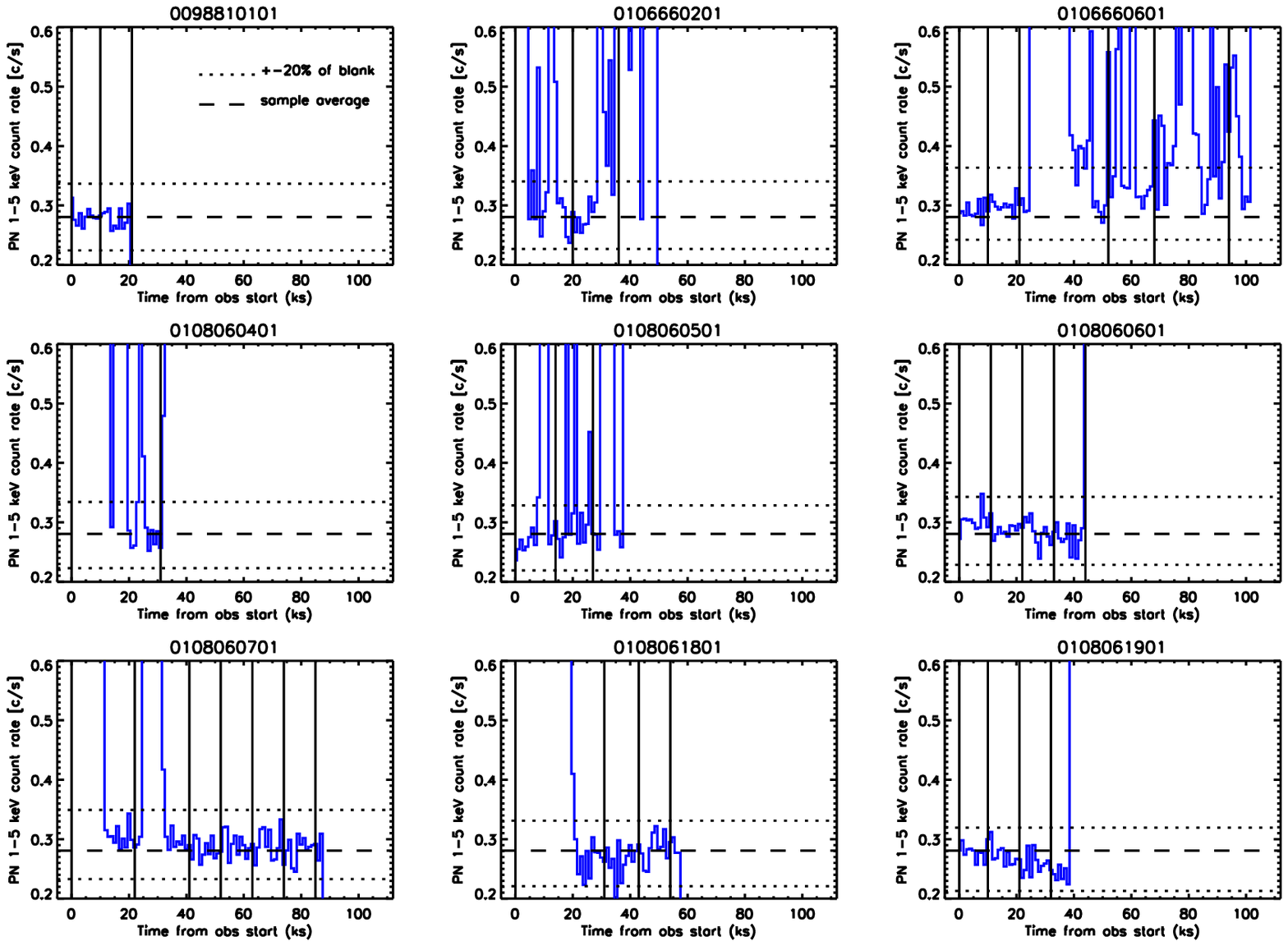}
\plotone{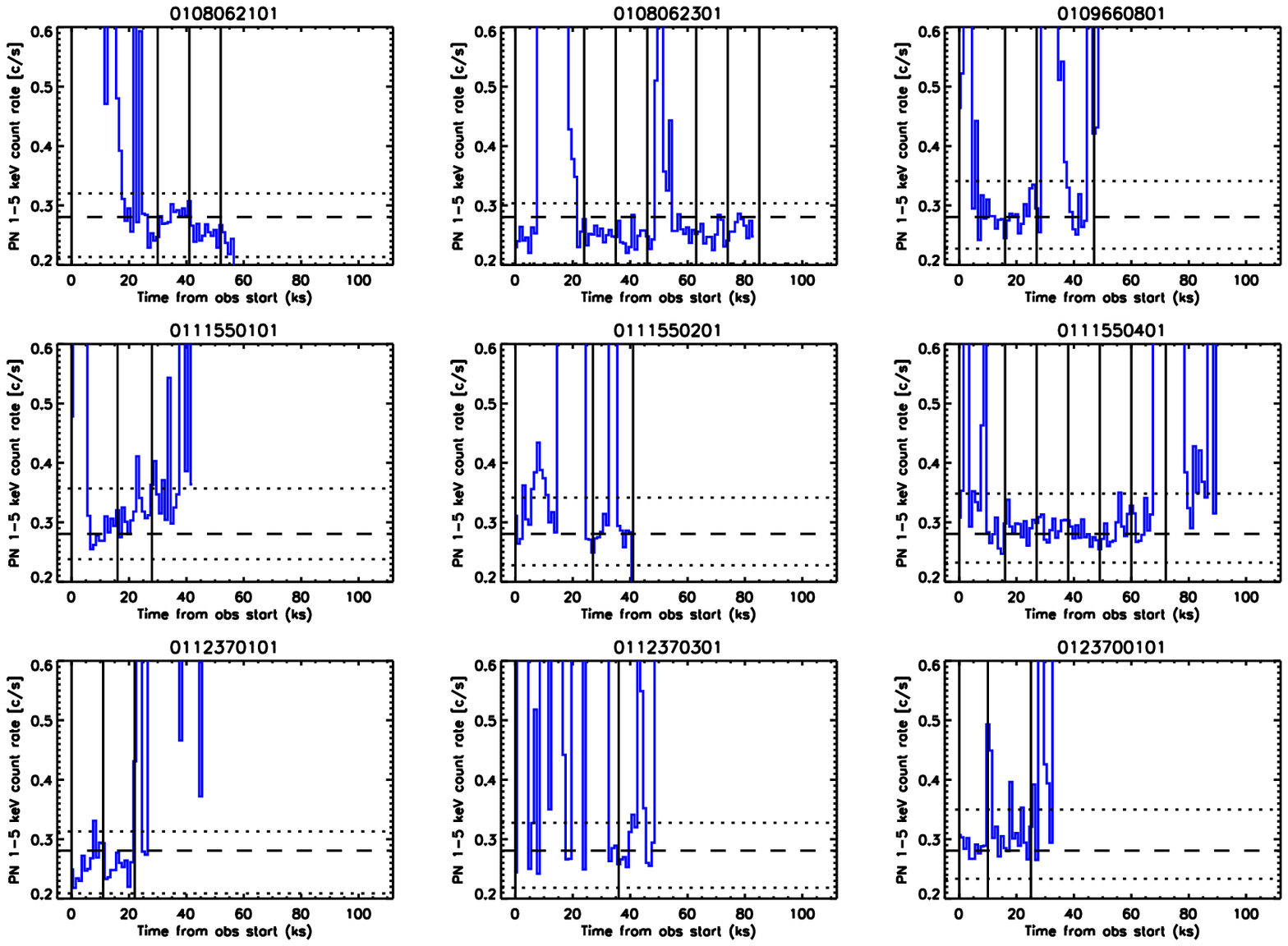}
\caption[]{PN 1--5 keV light curves in 1 ks bins in the 12--15 arcmin annuli
  for the observations listed in Table \ref{t2.tab}, together with the
  sample-average quiescent rate (dashed line) and the $\pm$ 20\% limits
  around the individual mean (dotted line).  The vertical lines show the
  start and stop times of the 10 ks periods used in the analysis.}
\label{f8.fig}
\end{figure}

Using our blank sky sample, we experimented with extracting light curves in
different soft bands in the 12--15 arcmin annulus.  We found that the 1--5
keV band is optimal for maximizing the number of counts in the 1 ks light
curve bins while minimizing the scatter of the resulting filtered
spectra. The 1--5 keV light curves are shown in Fig.\ 
\ref{f8.fig}. Similarly to the count rates in the hard band
(\S\ref{hardlc}), the resulting soft-band rate distribution of the
whole sample has a well-defined peak at low values.  
We chose to define the soft-band quiescent level as the 
mean (not exposure corrected, see \S\ref{anal}) count rate in the faint end 
ranges of 0.220--0.340 cts s$^{-1}$ for PN and 0.100--0.180 cts s$^{-1}$ for MOS1 and MOS2
for each pointing.
The average count rate of all pointings
in the above ranges are 0.281 cts s$^{-1}$ (PN) and 0.140 cts s$^{-1}$ (MOS1 and MOS2).
Using the $\pm$20\% limits around the mean in the above count rate range for
each pointing, we determined the soft band filtered quiescent periods
GTI$_s$ (see Table \ref{t3.tab}).  Note
that when applying the soft-band filtering to actual astronomical objects,
the absolute value of the quiescent level in the soft band may of course
include the source contribution. In order to achieve similar background
accuracy as in this work, one should use the $\pm$20\% limits of the absolute values of
our blank sky average, i.e.   $\pm$0.056 cts s$^{-1}$ (PN) and  $\pm$0.028 (MOS1 and MOS2) cts
s$^{-1}$ around the (variable) quiescent level of a given object, when
selecting quiescent periods (Appendix A).

Fig. \ref{f5.fig} shows that the soft-band filtering is also very
efficient in reducing the scatter in the hard band, which raises the
possibility of using only on the soft-band light curve. However, there is
residual hard-band variability due to non-simultaneous hard and
soft flares. Thus, to remove all detectable flares, one has to apply both
the $\pm$20\% hard-band filter and the $\pm$20\% soft-band filter (see
Appendix A), which we define here as the ``double-filtering'' method.
  Note that the addition of the soft band filter results in only
a 5--10\% reduction of the clean exposure time, compared to that
obtained with only the hard band filter (from 65\% to 55\% for PN and from
70\% to 65\% for MOS, see Table \ref{t3.tab}).

\section{Constructing the total quiescent background model}
\label{totmod}

We use here the double-filtered quiescent blank
sky spectra, derived above, for a detailed investigation of the background
modeling.

\subsection{Count rate correlations}
\label{corr}

In order to construct a background model based on the blank sky data, we
need to establish i) the stability of the spectral shape of the total
background and ii) a correlation between the count rates in the energy band
used for the scientific analysis, and the background level indicator
available in any extended source observation. The commonly used
indicator is the count rate at the highest energies, where the source
emission is eliminated due to negligible effective area.  To achieve the
above goals, we divided the double-filtered quiescent time periods for PN
and MOS into 50--70 pieces each about 10 ks long, similar to the typical
XMM-Newton observations. We extracted spectra obtained during the above
periods, within the full FOV, and grouped the spectra into
0.3--0.8--1.0--2.0--4.0--7.0 keV energy bins (and the hard-band bins
of 10--14 keV for PN and 9.5--12 keV for MOS), co-adding the MOS1
and MOS2 count rates to obtain spectra for MOS. These spectra have
been compared to see if the background variations in different energy
intervals are correlated.

\subsubsection{2--7 keV band}
\label{bias}

\begin{figure}
\plotone{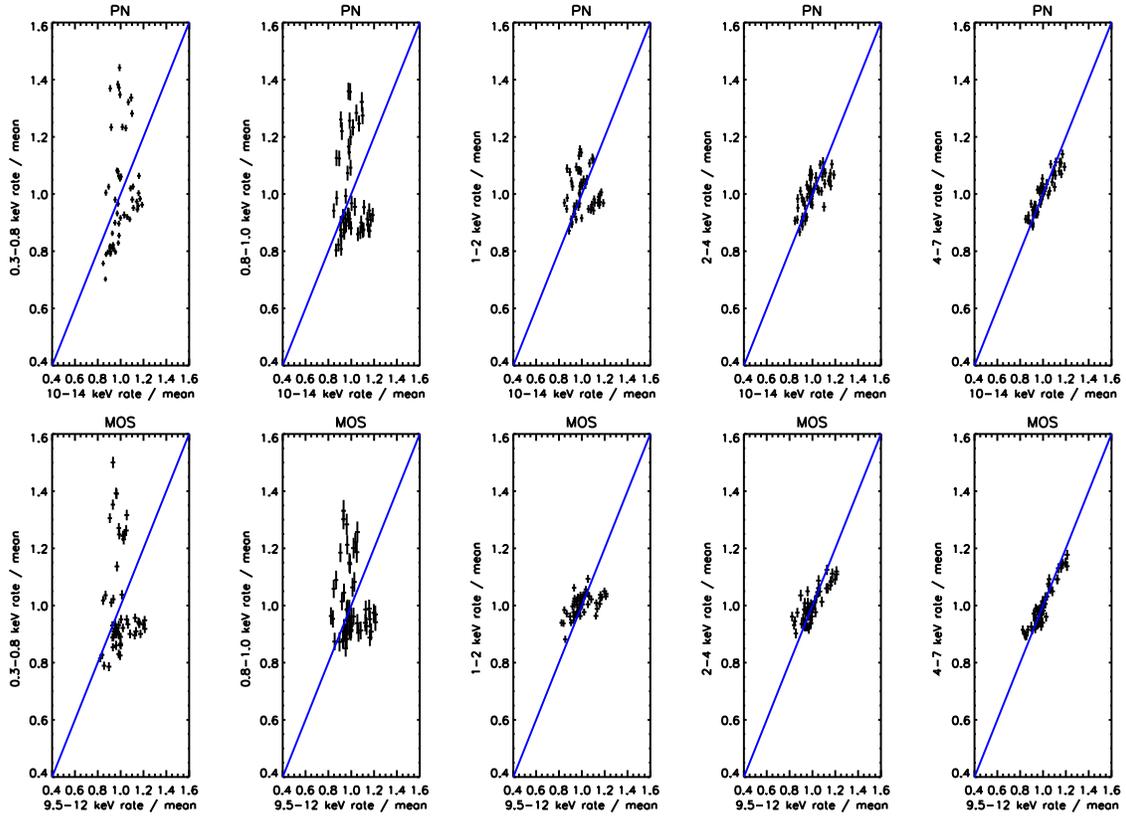}
\caption[] {Quiescent double-filtered PN (upper row) and MOS (lower row)
  count rates, divided by the mean values, in the full FOV region for the 10
  ks spectra sample, in 0.3--0.8--1--2--4--7 keV bins (columns) and in the
  hard band (10--14 keV for PN, 9.5--12 keV for MOS, rows) are shown by
  crosses. Uncertainties are statistical $1\sigma$. The lines show
  predicted rates assuming that the individual count rates in the intervals
  of the 0.3--7 keV band deviate from the average by the same factor as in
  the hard band.}
\label{f9.fig}
\end{figure}

Figure \ref{f9.fig} shows rates in our different 10 ks
``observations'' vs.\ the hard-band rate. Comparison of the data shows
(see Fig. \ref{f9.fig}) that the count rates correlate better with
the hard band rates at higher energies: the linear Pearson correlation coefficients (Neter et al. 1988) of the
2--4 and 4--7 keV rates with the hard band rates are 70\% and 90\% for PN,
85\% and 95\% for MOS, roughly consistent with Katayama et al.\ (2004).
When the hard-band rate does not exceed the mean value by more than 10\%,
the 2--4 and 4--7 keV rates deviate from the hard-band based linear
prediction (simple renormalization by the hard-band rate)
with a standard deviation of $\sim$ 5\%.  This is similar to the scatter of
the renormalized closed-cover spectra, implying that in this count rate
range there is no significant excess background component in addition to the
quiescent particle background.  However, when the hard-band rate exceeds the
average by more than 10\%, the rates in the 2--7 keV band deviate
downwards from the linear expectations (see Fig \ref{f9.fig}).
This suggests a residual particle background with harder spectrum than
that induced by the cosmic rays, perhaps due to the mild flares of the hard kind
mentioned in \S4.3, at a level below our filtering threshold.

\subsubsection{0.3--2.0 keV band}
\label{lowen}

Below 2 keV, the count rates deviate more strongly from the linear
expectation, reaching a maximum deviation of 50\%, and a standard
deviation of $>20$\%, at 0.3--0.8 keV.  The standard deviation of the
rates around the linear expectation significantly exceeds that of the
closed-cover sample at energies below 1.0 keV.
Note that the variation of $N_H$ in the sample (0.6 -- 3.0 $\times$
$10^{20}$ cm$^{-2}$) yields only a 5\% difference in the sky background
(i.e. Galactic + extragalactic photon background) flux at 0.8--1.0 keV, and
thus a negligible variation in the total background (see Fig.
\ref{f2.fig}).

Based on the RASS count rates in the fields of the XMM-Newton pointings in
our sample (see Appendix B), the spatially-variable Galactic
background at these energies does not explain all of the above variation.
This implies the existence of a time-variable background component
below 1 keV, possibly due to another particle population and/or
solar wind charge exchange background (indeed observed by {\em Chandra},
Wargelin et al.\ 2004). We did not find a practical way of incorporating 
the information on the spatio-temporal variability into the background modeling,
and thus we will not go into more detail here. 

\begin{deluxetable*}{lllllllllllll}
\tabletypesize{\scriptsize}
\tablecaption{Background accuracy}
\tablewidth{0pt}
\tablehead{
        & \multicolumn{4}{c}{hard band $\pm$ 50\% filter} & \multicolumn{4}{c}{hard band $\pm$ 20\% filter} & \multicolumn{4}{c}{double filter} \\
        &  \multicolumn{4}{l}{\hrulefill}   &  \multicolumn{4}{l}{\hrulefill}  &  \multicolumn{4}{l}{\hrulefill} \\
R       &  \multicolumn{4}{c}{band (keV)} &  \multicolumn{4}{c}{band (keV)} &  \multicolumn{4}{c}{band (keV)} \\
arcmin  & 0.8--1.0 & 1.0--2.0 &  2.0--4.0 &  4.0--7.0 & 0.8--1.0 & 1.0--2.0 &  2.0--4.0 &  4.0--7.0 & 0.8--1.0 & 1.0--2.0 &  2.0--4.0 &  4.0--7.0}
\startdata 
\multicolumn{12}{c}{{\bf PN}} \\
0--10  & 46 & 32 & 28 & 16 & 31 & 18 & 17 & 12 & 24 & 12 & 10 &  7 \\
10--15 & 39 & 28 & 23 & 16 & 29 & 19 & 17 & 13 & 20 & 11 & 8  &  5 \\
0--15  & 42 & 30 & 26 & 16 & 30 & 18 & 16 & 12 & 22 & 12 & 8  &  6 \\
       & & & & &    & & & & & & &  \\
\multicolumn{12}{c}{{\bf MOS}} \\
0--10  & 35 & 18 & 22 & 10 & 22 & 9  & 16 & 7 & 20 & 7 & 12 & 7  \\
10--15 & 29 & 15 & 20 & 10 & 21 & 9  & 9  & 6 & 17 & 7 & 7  & 7 \\
0--15  & 29 & 17 & 21 & 10 & 20 & 7  & 13 & 6 & 20 & 7 & 10 & 5 \\
       & & & & & & & & & & & &  \\
\enddata
\tablecomments{90\% confidence limits (\%) for the relative background
uncertainties, (prediction - data)/data, using the blank-sky sample.  The
columns ``hard band $\pm$50\% filter'' show values obtained using the
$\pm$50\% limits in the hard band around the nominal level to filter the
quiescent data, which approximates the method used in the literature. ``hard
band $\pm$20\% filter'' refers to filtering the blank sky data with
$\pm$20\% limits around the quiescent level in the hard band.  ``double
filter'' gives the values from our preferred method (see text).  For the
``hard band $\pm$50\% filter'', the simple scaling of the background model
is used, while for the ``hard band $\pm$20\% filter'' and ``double filter'',
the 0.8--2.0 keV values are obtained without any renormalization and the
2.0--7.0 keV values are obtained with a maximum of 10\% renormalization (see
\S\ref{method}).}
\label{t4.tab}
\end{deluxetable*}

\subsection{Method}
\label{method}
We use the hard band count rate to renormalize the blank sky spectrum to
correspond to a given observation, unless the computed renormalization
factor exceeds 1.1. In this case, we minimize the bias explained in \S\ref{bias} by using a maximum renormalization factor of 1.1.

Below 2 keV the count rates correlate very poorly with the hard band 
count rates, and thus a more detailed modeling would be favorable.  However, since 
many of the
deviations are below 10\%, the spectrum of the excess on top of the blank
sky based prediction is of too poor quality for proper spectral modeling,
and we do not attempt it here. Rather, we exclude the energies below 0.8
keV from further analysis in order to minimize the complexity of the
modeling, while maintaining enough counts for spatially resolved spectral
analysis.  We thus adopt the usage of the co-added blank sky spectrum for
the 0.8--2.0 keV band as well, but without any renormalization.  We will
examine the effect of the above choices on the background prediction
accuracy in \S\ref{accuquies}.

In the actual extended source analysis we extract the blank sky spectra
using the same detector region as used for extracting the source spectra
(see Appendix A).  The number of counts are often low in the hard band
within a small region.  Thus, to avoid statistical uncertainties, we further
assume that the ratio of the observed hard band count rate to that of the
blank sky spectrum is constant all across the detector, i.e. that the
renormalization factor as derived from full FOV region of the pointing in
question and the blank sky event file applies at all radii.

\section{Accuracy of the quiescent background prediction}
\label{accuquies}

For each blank sky observation, we form here a background model
prediction in the 0.8--7.0 keV band following the recipe described in
\S\ref{method}.  By comparison of the background prediction with the
observed blank sky spectrum, we evaluate the accuracy of the background
modeling.

\subsection{Individual pointings}

\begin{figure}
\epsscale{0.90}
\plotone{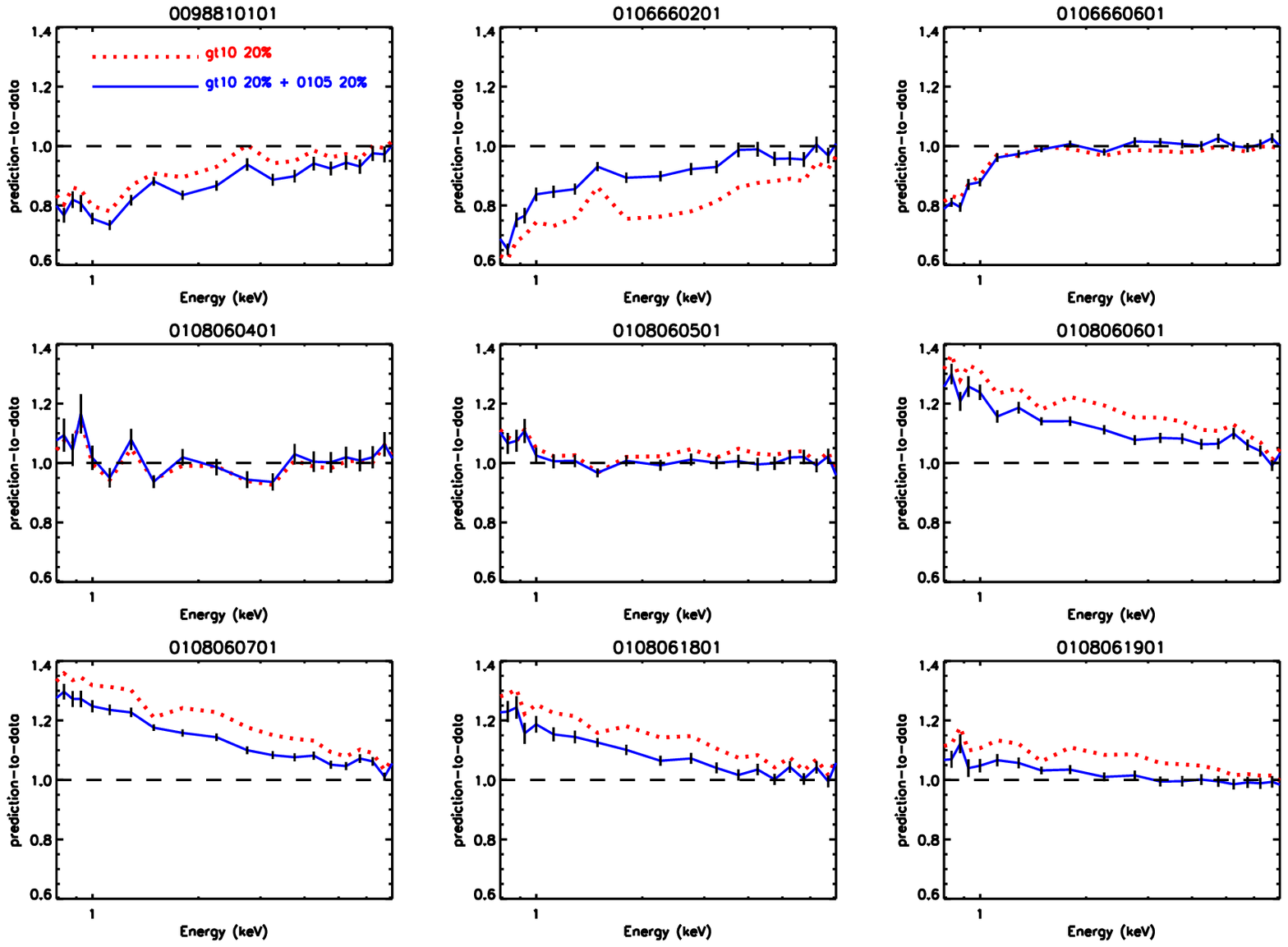}
\plotone{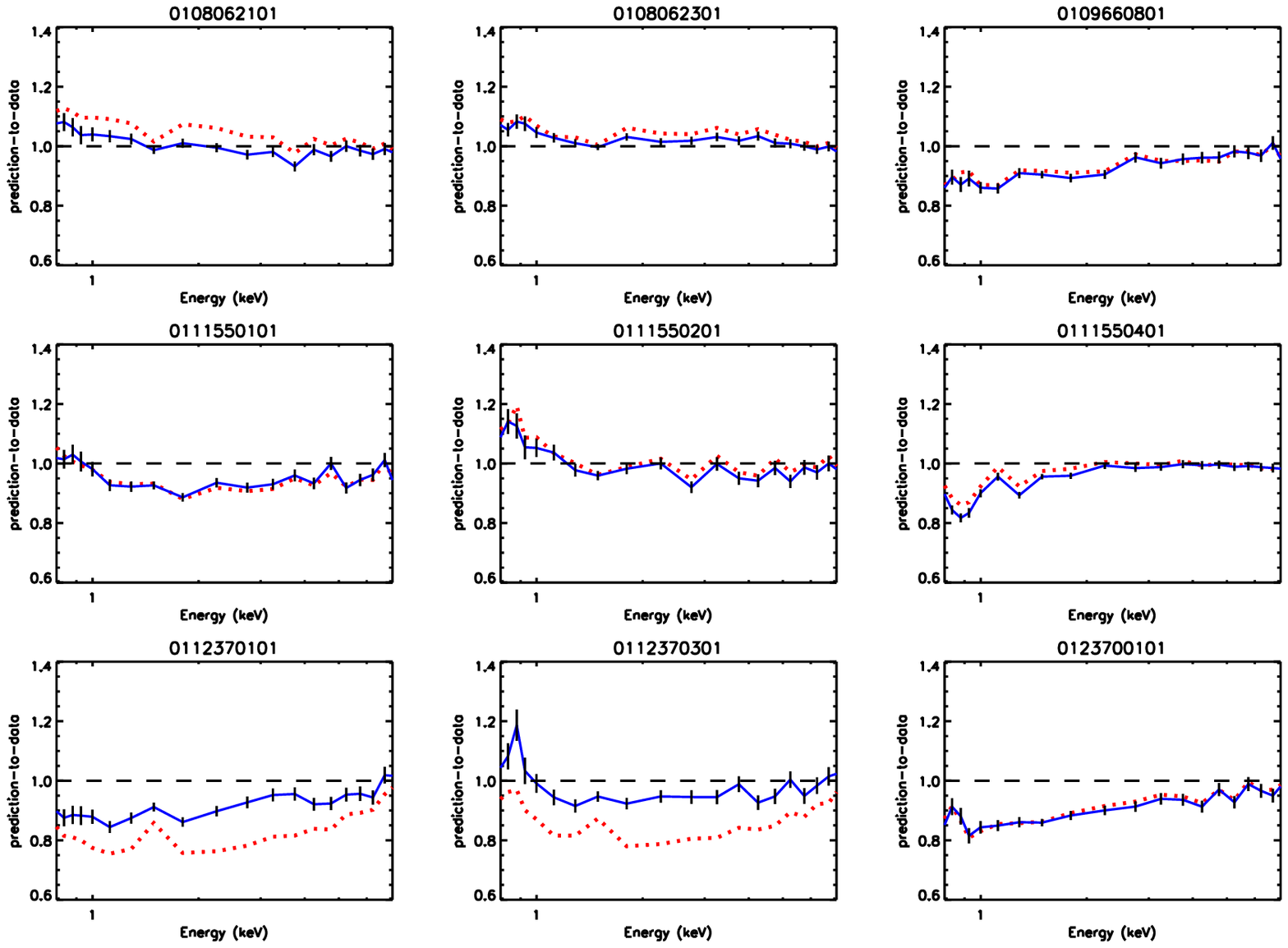}
\caption[]{The ratio of the background prediction rates to the observed
  rates for each of the individual blank-sky pointings, using only the hard
  band filter (dotted line) or the double filter (solid line). The vertical bars on the solid lines show the statistical 1 $\sigma$ uncertainties}
 \label{f10.fig}
\end{figure}

We first extracted the full FOV spectra of the blank sky pointings (see
Table \ref{t2.tab}) to maximize the signal-to-noise ratio in small
spectral bins.  For simplicity, we renormalized the full band (0.8--7.0 keV)
spectra using the hard band count rates (instead of using the maximum factor
of 1.1 at 2--7 keV band, and no renormalization at 0.8--2.0 keV).  The
comparison of the PN and MOS background spectrum prediction of the
individual pointings to the data reveals, that the background prediction of
the hard-band-filtered spectra is consistent with the statistical
uncertainties of 5\% at 7.0 keV (see Fig.  \ref{f10.fig} for
PN).  The maximum inaccuracy increases towards lower energies, reaching 20\%
at 1.0 keV, and 40\% at 0.8 keV, demonstrating the need for more complex
modeling at lowest energies.  The usage of the double filter (see \S\ref{softfil}) instead, improves the accuracy systematically (see Fig.
\ref{f10.fig}).  The accuracy of MOS1 and MOS2 is very
similar, and thus in the following spectral analysis we use co-added MOS1
and MOS2 blank sky data to obtain results for MOS.

\subsection{10 ks spectra}

In the following we examine the background accuracy with better temporal and
radial resolution, in order to derive information more applicable to actual
extended source analyzes.  For the double-filtered data, we use the 10ks
periods already introduced in \S\ref{corr}, and we determine
corresponding 10ks periods for the hard-band-filtered data.  We extracted
blank sky spectra in 0--10 and 10--15 arcmin annuli, around the center
of FOV.  
We binned the spectra in the bins of 0.8--1.0--2.0--4.0--7.0 keV.  We
use the maximum renormalization factor of 1.1 in the 2--7 keV band, and no
renormalization in the 0.8--2.0 keV band, as explained in \S\ref{method}.

\subsection{Results}
\subsubsection{Hard band vs.double filtering}

Comparison of the 10 ks PN spectra with the predictions using the hard band
filter shows that the data deviate most from the prediction (by +60\%) in
the bin 1 of pointing 0106660201 and in the bin 3 of pointing 0112370101
(Fig. \ref{f3.fig}).  In these periods there are some instants
when the soft band count rate is elevated while the hard band appears
consistent with the quiescent level. Thus the addition of the soft band
filter cleans the data very efficiently resulting in 30\% maximum relative
uncertainties at 0.8--1.0 keV band.  The hard-band-filtered MOS data are
overall in better agreement with the prediction and the addition of the
soft band filter makes a smaller, but still systematic improvement (see
Table \ref{t4.tab}).

We used the distribution of the data-to-model ratio for each energy to
determine the 90\% confidence limits for the background accuracy.  We
performed the calculation to both double-filtered, and hard-band-filtered
data.  Conservatively, we report the bigger of the positive and negative
deviations. Considering the full FOV spectra, the addition of the soft
band filtering improves the PN accuracy, compared to hard-band-only
filtering, from 10--30\% to 5--20\% in the 0.8--7 keV band (see Table
\ref{t4.tab}) at 90\% CL, while MOS accuracy improves less significantly,
staying at 5--20\%.  The accuracy of the double-filtered full FOV
background of PN and MOS improves with increasing energy, from 20\% at
0.8--1.0 keV to 5\% at 4--7 keV. The accuracy improves systematically, but
marginally, with increasing radius, suggesting that the residual flares
are brighter in the middle of the FOV.

\subsubsection{Uncertainty correlations}
\label{errcorr}

\begin{figure}
\epsscale{0.90}
\plotone{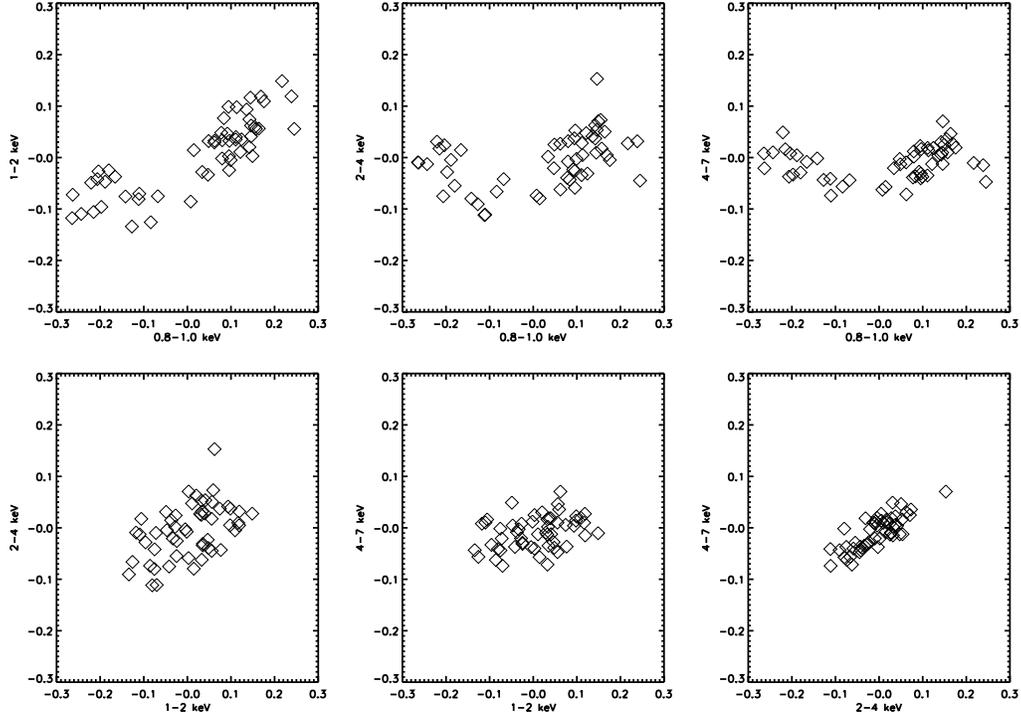}
\caption[]{Relative deviations of the observed PN backround from our
  predictions in different bands, (prediction - data)/data. The data are
  cleaned using the double-filter method and binned into 10 ks time bins.}
\label{f11.fig}
\end{figure}

As shown above, the variable residual background components give
different contribution in different bands, and thus the model
uncertainties in the different bands are not strongly correlated (see
Fig.  \ref{f11.fig}). However, at the highest energies the
correlations are stronger than at the lowest energies, implying an
increasing complexity of the background modeling towards lower energies (as
noted in \S\ref{lowen}, see also Appendix B).

To propagate accurately the model uncertainties to those of the final
spectral fits (e.g., cluster temperatures), one needs to take into account
these correlations. The most conservative approach would be to assume no
correlation between the different wide bands and simply vary the
normalizations of the background model in different bands by their
90\% confidence level values (see Table \ref{t4.tab}), when evaluating the
effect of the background uncertainties. This would result in extreme
situations, e.g., with the E $<$2 keV background at upper (lower) 90\%
CL limit, while the E $>$2 keV background at lower (upper) 90\% CL limit.
Such situations would yield maximum variation in the temperature
measurement, but fortunately the uncertainty correlations rule out such
extremes (see Fig.\ \ref{f11.fig}).  Thus, the inclusion of the
uncertainty correlation information would result in smaller (and more
correct) uncertainties in the resulting temperature measurements.  Due to
the large scatter and weak correlations, it is not practical to apply
these results analytically to spatially resolved spectroscopy.  Rather, in
\S\ref{a1795} we will describe a practical method to apply this
information numerically.

\subsubsection{Renormalization choices}

\begin{deluxetable}{lllll}
\tablecaption{Full FOV 90\% CL accuracy}
\tablewidth{0pt}
\tablehead{
method  &  \multicolumn{4}{c}{band (keV)} \\
        & 0.8--1.0 & 1.0--2.0 &  2.0--4.0 &  4.0--7.0 }
\startdata 
\multicolumn{5}{c}{{\bf PN}} \\
no separ., simple  & 29 & 18 & 11 & 6 \\
no separ., impr    & 22 & 12 & 8  & 6 \\
separ.   , impr    & 22 & 12 & 8  & 5  \\
\multicolumn{5}{c}{{\bf MOS}} \\
no separ, simple    & 25 & 15 & 10 & 6 \\
no separ, impr      & 20 & 7  & 10 & 5 \\
separ., impr        & 20 & 7  & 7  & 5  \\
\enddata
\tablecomments{90\% confidence limits (\%) for the relative background
uncertainties for full FOV region obtained with different renormalization
methods and the double-filtered blank sky data.  ``separ'' and ``no separ''
refer to separating or not the particle background before renormalization.
``simple'' refers to renormalizing the full band with the same factor, and
``impr'' refers to the method or using maximum factor of 1.1 in 2--7 keV
band, and no renormalization in the 0.8--2 keV band.}
\label{t5.tab}
\end{deluxetable}

We examine here the effect of our adopted choices of the maximum
renormalization factor of 1.1 in the 2--7 keV band and no renormalization in
the 0.8--2.0 keV band, on the accuracy of the prediction of the full FOV
region background spectrum.  For comparison of the resulting uncertainties
with the above choices (derived above, see Table \ref{t4.tab}), we computed
the uncertainties simply by renormalizing the co-added double-filtered blank
sky spectrum in the full 0.8--7.0 keV band based on the hard band rate.  The
comparison shows (see Table \ref{t5.tab}) that our choice improves the
modeling significantly, most importantly at energies below 2 keV.

\subsubsection{Separating the particle background}

When renormalizing the total co-added blank sky spectra with the hard band
rate (consisting only of particle background), we effectively assume that
the sky background spectrum of a given observation deviates from the sample
average by the same factor as the particle background of a given observation
deviates from the sample average. However, the sky background and the
particle background are independent, and thus we would introduce a bias
by the simple renormalization method. However, if we apply the
renormalization only in the 2--7 keV band, where the sky background
contributes only $\sim$20\%, the effect of the bias should be
negligible.  To verify this, we computed the uncertainties of the
soft-band-filtered background model, separating first the quiescent particle
background (estimated from the closed-cover spectrum), renormalizing it
by the hard-band count rate, and adding it back to the remaining sky
background model.  As expected, the resulting uncertainty values (see Table
\ref{t5.tab}) for the full FOV are better than those for simple
rescaling.  However, they are not significantly different from those
obtained with our adopted scaling method (see \S\ref{method}), which
is much more practical.

\section{Application to A1795}
\label{a1795}

To illustrate the application of the proposed background modeling
procedure (See Appendix A), we analyzed the XMM-Newton observation
0097820101 of the bright nearby cluster of galaxies A1795, previously
analyzed by Arnaud et al.\ (2001) and Nevalainen et al.\ (2002).  The
cluster was observed on 2000 June 26, with thin filters and in Full
frame readout mode.  We concentrate on regions outside the bright cool core (r$>$ 3
arcmin) where the details of the background modeling are important.  We
processed the A1795 data using the same procedure as the blank sky data
(\S\ref{anal}).  The LIVETIME values for different CCDs in a given
instrument vary by less than 1\%.  At high energies, CCD 7 of MOS2 has a
significantly higher count rate level, compared to other CCDs, so we
excluded it.  We further filtered the event files with expression
``flag==0'', and including only patterns 0--4 (PN) and 0--12 (MOS).  In
0.8--7 keV band, $epatplot$ tool in SAS package shows that PN and MOS single
and double pixel event distributions agree with the model used when generating
the detector response matrices (Ehle et al., 2004).

\newpage

\subsection{Light curves}
\label{a1795lc}

\begin{figure}
\epsscale{0.90}
\plotone{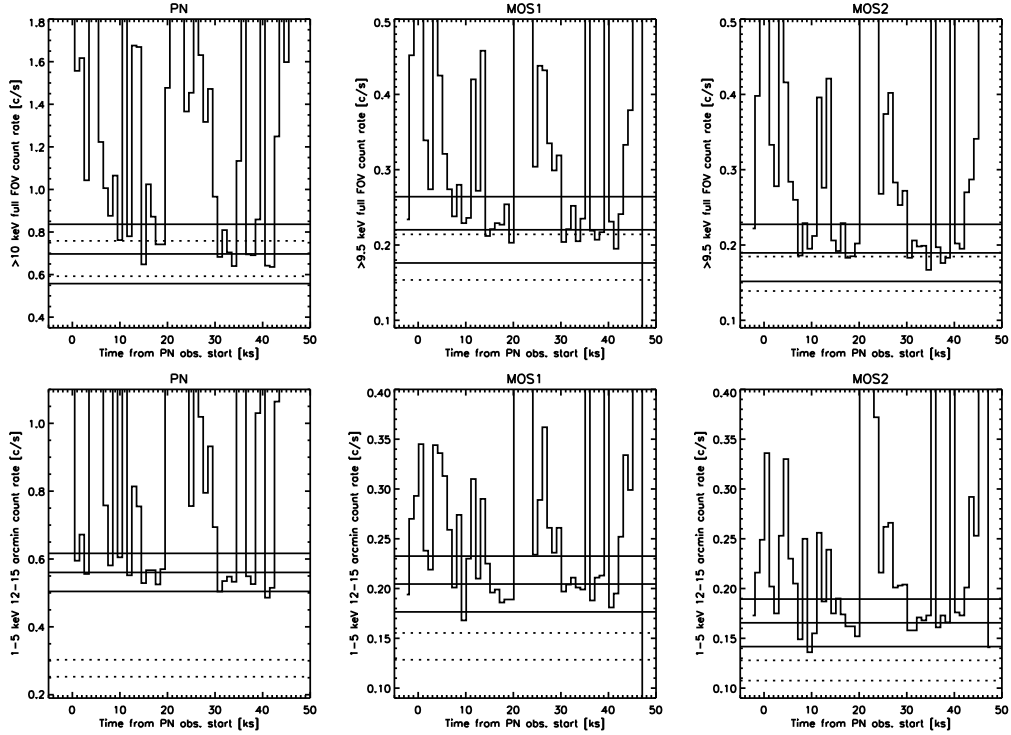}
\caption[]{A1795 light curves (histograms) in the hard band for the the
  full FOV (upper row) and in the 1--5 keV band for the 12--15 arcmin
  annulus (lower row), for PN (left colum), MOS1 (middle column) and MOS2
  (right column).  The quiescence limits are shown with a solid line; the
  highest and the lowest of the individual blank-sky quiescence levels (see
  Table \ref{t3.tab}) are shown with dotted lines. The MOS2 blank-sky
  values are reduced by 15\% due to the excluded peculiar CCD (see text).}
\label{f12.fig}
\end{figure}

Following the methods presented above, we extracted the PN and MOS light
curves of A1795 in 1ks time bins in the hard band ($>$ 10 keV for PN, $>$
9.5 keV for MOS) within the full FOV, and in the soft band (1--5 keV) within
the 12--15 arcmin annuli, excising point sources (see Fig.
\ref{f12.fig}).  Consistently with the blank sky analysis, we do not
apply the out-of-time correction at this stage to PN data (it will be
applied later).  Also, we report here the count rates as obtained from the
light curves without applying an exposure correction (see Table
\ref{t6.tab}), as in the case of the blank sky data (see \S\ref{anal}).

\begin{deluxetable}{llllll}
\tablecaption{A1795 quiescence}
\tablewidth{0pt}
\tablehead{
\colhead{Instr.}   & \multicolumn{2}{c}{hard}       & \colhead{soft} & \multicolumn{2}{c}{double}      \\
                   & \multicolumn{2}{l}{\hrulefill} &                & \multicolumn{2}{l}{\hrulefill}  \\
                   & \colhead{cr} & \colhead{tfrac} & 
                     \colhead{cr} & \colhead{tfrac} &  \colhead{exp} \\
                   & \colhead{[cts s$^{-1}$]} & \colhead{[\%]} & \colhead{[cts s$^{-1}$]} & \colhead{[\%]} & \colhead{[ks]} }
\startdata 
PN   & 0.697       & 28  & 0.560              & 23  & 10 \\
MOS1 & 0.220       & 42  & 0.205             & 38  & 19 \\
MOS2 & 0.190       & 38  & 0.166             & 34  & 17 \\
\enddata
\tablecomments{``cr'' shows the mean count rates of A1795 in the full FOV in the hard band (``hard'') and in the 12--15 arcmin annulus in the soft band
(``soft'') in the faint end, i.e. the quiescent levels (see \S4.2 and \S4.4). 
MOS2 values are obtained excluding the spoiled CCD 7.  The count rates are obtained using the light curves of 1 ks time bin, without applying exposure correction. 
``tfrac'' shows the fraction of the selected good time intervals, compared to total, using the hard band light curves, 
and when applying the double filter (``double''). ``exp'' shows the resulting exposure time (including exposure correction) when applying the 
double filter}
\label{t6.tab}
\end{deluxetable}

In the hard band, using the same count rate range for averaging as for
the blank sky (0.54--0.80 cts s$^{-1}$ for PN and 0.12--0.24 s$^{-1}$ for
MOS1, see \S\ref{hardlc}) we obtain quiescent levels of 0.697 s$^{-1}$
(PN) and 0.220 s$^{-1}$ (MOS1). Due to the exclusion of CCD 7 of MOS2, we
lowered the nominal MOS2 limits by 15\%, obtaining a quiescent rate of 0.190
cts s$^{-1}$.  These are consistent with the largest deviation observed in the
blank-sky sample.  Using only periods when the count rate is within
$\pm$20\% of the above levels, we determined the quiescent periods GTI$_h$.
The hard band filter selects 30\% (PN) and 40\% (MOS) of the total exposure
time. 

In the 1--5 keV band in the 12--15 arcmin annulus of PN the cluster
contribution is $\sim$ 50\% of the total A1795 count rate, while due to the
smaller effective area the corresponding fraction is 30\% in MOS1.  Using
the faint end count rate distributions (0.40--0.70 cts s$^{-1}$ (PN);
0.15--0.25 cts s$^{-1}$ (MOS1); 0.10--0.20 cts s$^{-1}$ (MOS2) we obtained the
average quiescent (cluster + background) levels as 0.560 cts s$^{-1}$ (PN),
0.205 cts s$^{-1}$ (MOS1), 0.166 (MOS2).  We used the absolute values of
$\pm$20\% of the blank sky average ($\pm$0.057 cts s$^{-1}$ for PN and
$\pm$0.028 cts s$^{-1}$ for MOS1, see \S\ref{softfil}) around the
above quiescent levels as the range to determine the quiescent periods
GTI$_s$, except for MOS2, for which we used 15\% smaller limits ($\pm$0.023 cts
s$^{-1}$) due to the exclusion of CDD 7.  The useful exposure times using
the double filter (i.e.  using the common accepted times in GTI$_h$ and
GTI$_s$) reduce to 25\%, 40\% and 35\% of the total for PN, MOS1, and MOS2.

Note that the application of only the soft band filter to PN data obtains 2
ks longer useful exposure than the application of only the hard band filter,
i.e. in the case of A1795 the soft band filter is less effective than the
hard band filter, in cleaning the flares. This is due to the presence of
flares with hard spectrum (see Fig. \ref{f12.fig}), which
illustrates the need for double filtering.

\subsection{Spectral analysis} 
\label{a1795quies}

\begin{deluxetable}{llll}
\tablecaption{Double-filtered count rates in the hard band}
\tablewidth{0pt}
\tablehead{
\colhead{}  & \colhead{A1795}      & \colhead{$<$blank$>$}  & \colhead{renorm}  \\
\colhead{}  & \colhead{cts s$^{-1}$} & \colhead{cts s$^{-1}$}   & \colhead{}}
\startdata 
PN   & 0.499$\pm$0.007 & 0.453$\pm$0.001 & 1.10$\pm$0.03 \\
MOS  & 0.117$\pm$0.004 & 0.113$\pm$0.000 & 1.03$\pm$0.03 \\
\enddata
 \tablecomments{Count rates and 90\% CL statistical uncertainties of A1795 and the blank sky ($<$blank$>$) in the 10--14 keV (PN) and 9.5--12 keV (MOS) bands for the double-filtered quiescent data in the full FOV, except that CCD 7 of MOS2 is excluded, and the renormalization factors (renorm). The values are obtained from the exposure-corrected spectra}
\label{t7.tab}
\end{deluxetable}

We extracted the quiescent PN and MOS spectra in concentric annuli of radii
3--4--5--7--10--15 arcmin and in a full FOV circle, using the double filter
(GTI$_s$ and GTI$_h$, obtained in \S\ref{softfil} and \ref{hardlc}),
excising point sources. We converted our background event files into sky
coordinates of the A1795 observation and used the source region files in sky
coordinates to extract the blank-sky spectra. 
The usage of the same region files minimized the differences in the size of the 
source and the background spectrum accumulation area below 1\% level in our case. 
We co-added the MOS1 and MOS2 spectra (excluding MOS2 CCD 7), and averaged
the responses.  The exposure times are 10 ks and 36 ks for PN and the
combined MOS spectra, respectively. Using the blank sky count rates of the double-filtered
spectra of the full FOV in 10--14 keV (PN) and 9.5--12 keV (MOS) bands, we
obtained the renormalization factors of 1.10$\pm$0.03 (PN) and 1.03$\pm$0.03 (MOS) at 90\% CL (see Table
\ref{t7.tab}). The uncertainties result from the statistical uncertainties of A1795 hard band
counts rates of 2\% (PN) and 3\% (MOS). In the case of long A1795 observation these small uncertainties
yield a negligible variation in the temperature measurements, compared to the other background modeling uncertainties
(Table 8). In any given observation, this uncertainty should be evaluated and if significant, it should be 
propagated to the temperature measurements.

\begin{deluxetable}{llllll}
\tabletypesize{\footnotesize}
\tablecaption{Temperatures}
\tablewidth{0pt}
\tablehead{
\colhead{R}       & \multicolumn{3}{c}{double filtered} & \multicolumn{2}{c}{hard band $\pm$50\%}        \\
\colhead{}        & \multicolumn{3}{l}{\hrulefill}      & \multicolumn{2}{l}{\hrulefill}  \\
\colhead{arcmin}  & \multicolumn{3}{c}{T [keV]} & \multicolumn{2}{c}{T [keV]} \\
\colhead{}        & \multicolumn{3}{l}{\hrulefill} & \multicolumn{2}{l}{\hrulefill}       \\
\colhead{}        & \colhead{stat} & \colhead{exact} &  \colhead{appr} & \colhead{stat} & \colhead{exact} }
\startdata
\multicolumn{6}{c}{\bf PN} \\
3--4    & 5.4[5.1--5.8] & 5.4--5.5 & 5.4--5.5  & 5.3[5.1--5.5] & 5.2--5.4 \\ 
4--5    & 5.0[4.7--5.5] & 5.0--5.1 & 5.0--5.1  & 5.2[4.9--5.5] & 5.0--5.3 \\
5--7    & 5.2[4.8--5.6] & 5.1--5.4 & 5.1--5.4  & 5.1[4.9--5.4] & 4.7--5.4 \\
7--10   & 4.7[4.3--5.3] & 4.4--5.1 & 4.4--5.1  & 5.2[4.8--5.6] & 4.0--6.1 \\
10--15  & 4.3[3.5--5.4] & 3.5--5.5 & 3.3--5.6  & 4.2[3.6--5.1] & 2.0--8.2 \\
        &               &          &           &               &         \\
\multicolumn{6}{c}{\bf MOS} \\
3--4    & 6.0[5.7--6.3] & 5.9--6.0 & 5.9--6.0  & 6.0[5.7--6.2] & 5.8--6.0 \\ 
4--5    & 6.0[5.6--6.4] & 5.9--6.1 & 5.9--6.1  & 5.9[5.7--6.2] & 5.7--6.0 \\
5--7    & 5.8[5.5--6.2] & 5.6--6.0 & 5.7--6.0  & 5.8[5.6--6.1] & 5.3--6.1 \\
7--10   & 5.6[5.1--6.2] & 5.0--6.0 & 5.1--6.2  & 5.8[5.4--6.3] & 4.4--6.8 \\
10--15  & 4.9[4.2--5.8] & 3.6--6.0 & 3.5--6.8  & 5.3[4.6--6.2] & 2.2--9.3 \\
        &               &          &           &               & \\
\enddata
 \tablecomments{
The PN and MOS temperature profile values of A1795 obtained using the double filtering method (``double filtered'') or using the approximated fixed $\sigma$ clipping 
method (``hard band $\pm$50\%''). 
``stat'' shows the best fit temperatures in the 0.8--7 keV band with statistical uncertainties. 
The variation of the best-fit temperature when including the background uncertainty correlation information using the exact method (\S 7.2.1)
is denoted as ``exact'', and the variation resulting from the approximate background uncertainty propagation method is denoted as ``appr'' (\S 7.2.2). 
Note that the systematic uncertainties ``exact'' and ``appr'' are independent of the statistical uncertainties in the column ``stat''. 
The uncertainties are reported at 90\% confidence level}
\label{t8.tab}
\end{deluxetable}

For PN, we generated the simulated out-of-time (OOT) event file as
described in Ehle et al.\ (2004).  We then subtracted the OOT spectra
normalized by a factor of 6.3\% (the ratio of CCD readout time to frame time
in Full frame mode) from the cluster spectra.  The subsequent
subtraction of a not-OOT-corrected blank-sky spectrum normalized as described above would
oversubtract the background, because the simulated OOT emission already
includes 6.3\% of it. Thus, before subtracting the blank-sky spectra, we
have reduced their normalization calculated above, by 6.3\% 
\footnote{The background consists of observations performed in Full frame and Extended full frame readout modes  
and thus the exact reduction factor should be derived by averaging the corresponding readout time fractions (6.3\% and 2.3\%) using exposure 
times as weights. However, we show in Appendix C that approximating the exact factor by 6.3\% 
results in negligible effect on A1795 temperature measurements}. 
This assumes that the background is spatially uniform so that randomization of
one of the coordinates during the generation of the OOT event file does not
change it. Given the small fraction of the OOT background, this is a
reasonable approximation (see Appendix C).

For spectral fits, we excluded the energy interval 1.45--1.55 keV which
contains time-variable Al K instrumental lines. 
Exclusion of this band is also necessary in order to avoid incorrect spatial localization 
of instrumental Al lines due to the not corrected OOT effect on the background data.
To propagate the different
background uncertainties at different energies, and to perform the
background model renormalization in an energy-dependent manner
(specifically, only in the 2--7 keV band), we divided each spectrum into 4
bands, 0.8--1--2--4--7 keV, and treated them as separate spectra with tied
model parameters (see Appendix A). We calculated the background
normalization taking into account the above renormalization factors 
and the OOT correction (see Appendix A). The PN and combined MOS spectra of
the cluster were fit with a MEKAL model and Galactic absorption with $N_H =
1.0 \times 10^{20}$ cm$^{-2}$. The resulting best-fit temperatures in
different radial bins are given in Table \ref{t8.tab} and in Fig.\ 13.
There is some evidence of multiple thermal components at the same radii
(as noted, e.g., by Nevalainen et al.\ 2003, when including the lowest
energies 0.3--0.8 keV in the fit). This may be part of the reason why
the PN best fit values are systematically lower than those of MOS (see Table
\ref{t8.tab}).  However, the single temperature model is adequate for
our present purpose of studying the background systematics.

\subsubsection{Exact propagation of the background uncertainties}
\label{mosbkg}
We will first try to include the background uncertainties into the
temperature fits using the full information on uncertainty correlations
obtained from our sample of blank-sky spectra (see \S\ref{errcorr}). Each 10 ks blank-sky spectrum corresponds to a different
realization of the background scatter in different bands, which we
use to modify the normalization of the background model in each
band. 
We fitted the A1795 spectra using the
background normalizations calculated from each of the 10 ks blank-sky
piece, and used the resulting distribution of the best-fit
temperatures to derive the systematic effect of the background
uncertainty for each radial bin.  

\begin{figure*}
\plotone{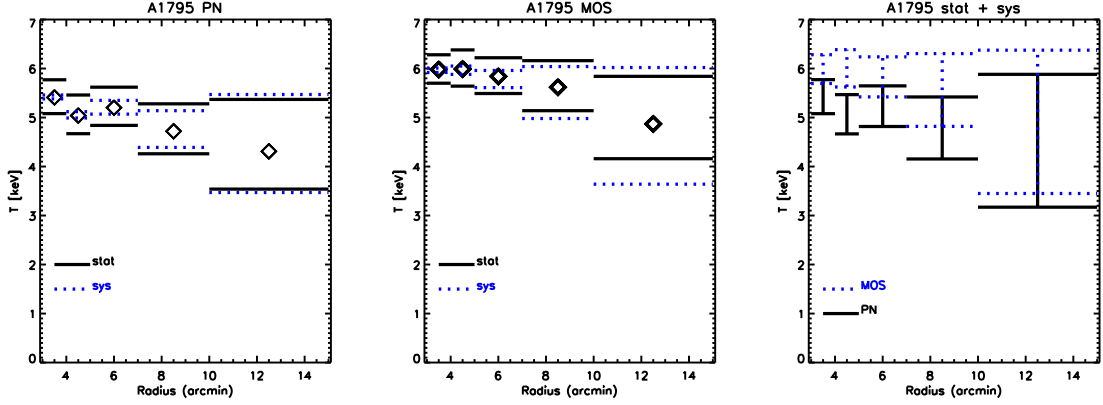}
\caption[]{Best-fit temperatures (diamonds) of A1795 in the
  $r=3-4-5-7-10-15'$ annuli for the double-filtered quiescent periods of PN
  (left panel) and MOS (middle panel), together with the statistical
  uncertainties (solid black lines, denoted as ``stat'') and the
  background-induced systematic uncertainties using the exact error propagation method (\S 7.2.1, dotted blue lines, denoted as
  ``sys'') at the 90\% confidence level. Right panel shows the combined
  effect of the statistical and the background-induced systematic
  uncertainties (added in quadrature) on the temperature profile for PN (black solid line) and MOS
  (dotted blue line).}
\label{f13.fig}
\end{figure*}

Due to its larger effective area, the cluster signal in PN  
is relatively stronger, compared to the background, than in MOS. Therefore the background uncertainties have a bigger 
effect on MOS than on PN (see Table 8).
The increasing fraction of the background with the larger radius increases
the background-induced uncertainty in the best-fit T from a few \% (90\% CL)
at $r=3-4'$ to $\sim$20--25\% (i.e., $\sim$ $\pm$1 keV for A1795)
at $r=10-15'$ (Fig.\ \ref{f13.fig}). 
Thus, even though the systematic
uncertainties exceed the statistical ones ($\sim$20\% ) at 10--15
arcmin, they do not prohibit the derivation of cluster temperature
profiles at $r>10'$ --- provided the flares are cleaned conservatively as
we do in this work.

\subsubsection{Approximate propagation of the uncertainties}

To reduce the complexity of the fitting procedure, we sought a simpler
error propagation method approximating the above effect on the temperatures.
We found that if we vary the background model simultaneously in the full
0.8--7.0 keV band by $\pm$10\%, the scatter of the best-fit temperatures
is quite close to that obtained using the full uncertainty
correlation information (see Table \ref{t8.tab}). The only significant
difference is the MOS temperature at 10--15 arcmin, whose 20\% uncertainty
is overestimated as 40\%. This may be due to the larger background-to-source flux ratio
in MOS than in PN (see Section \S\ref{mosbkg}). Basing only on one cluster we cannot estimate
how well the approximation works in different clusters. However, it is reasonable to assume
that the approximate method yields conservative estimates on the uncertainties.
Thus, if the approximate method yields uncertainties that are too large for the scientific
goals when analyzing a given object, one could perform the more complicated exact error propagation method,
to decide whether meaningful results can be derived for the object in question. 

\subsection{Comparison with flare cleaning from the literature}
\label{approxim}

It is useful to compare the temperature uncertainties obtained with our
method of flare filtering (double filter using the $\pm$20\% limits), with
those obtained with the less restrictive flare filtering method
commonly used in XMM-Newton cluster work (e.g., Pratt \& Arnaud
2002, 2003; Reiprich et al.\ 2003). To do this, we repeated our A1795
analysis using only the hard band for flare filtering and increasing the
light curve clipping limits to $\pm$50\% around the quiescent level (the
actual clipping limits used in those works are 50--100\%, see \S\ref{hardlc}, 
so our derived uncertainties will in fact be lower limits).
We also have re-created the blank-sky datasets, a combined one and one
divided into 10 ks pieces, using a similar, less restrictive flare
cleaning. We then used the hard-band rate to renormalize the background
model in the whole energy band and compared it with the 10 ks
background pieces to obtain the uncertainty distributions.  As expected,
the background scatter is significantly greater than in our method, because
of the inclusion of higher levels of the the spectrally variable flare
emission (see \S\ref{mildflare}). The scatter reaches $\pm$45\%
in the lowest energy bands (see Table \ref{t4.tab}).

\begin{figure*}
\plotone{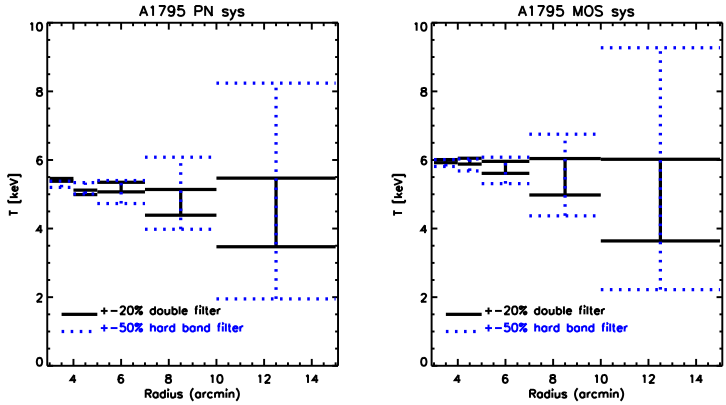}
\vspace{-5cm}
\caption[]{Background-induced systematic uncertainties using the exact error propagation method (\S 7.2.1) for the A1795
  temperatures in our double-filtering method are shown by solid black line,
  while those in the less restrictive flare cleaning common in the
  literature (see text) by blue dotted line. PN is shown in left panel and
  MOS in right panel.}
\label{f14.fig}
\end{figure*}

We then included the resulting background uncertainty correlations into
the A1795 spectral fits (\S7.2.1). The results are shown in
Fig.\ \ref{f14.fig}. At $r>5'$, the systematic temperature uncertainties
in the common method are 2--4 times as large as those in our 
method. At $r>10'$, these uncertainties become so large (50--100\% for PN,
60--80\% for MOS) that they render the temperatures essentially
unconstrained. This is despite the shorter clean cluster exposures that we
used. This comparison emphasizes the importance of the conservative flare
filtering when fitting spectra of low-brightness cluster regions.

\section{Conclusions}

We analyzed samples of XMM-Newton EPIC blank-sky and closed-cover data in order to explore the background subtraction methods for such extended sources 
which fill the whole EPIC FOV and thus render the local background estimation difficult. 
We found that during some quiescent periods, as determined by the hard band, the 1--5 keV band count rate reaches a factor of 3 times the 
quiescent level. Similarly, during some quiescent soft band periods, there are hard flares. Due to these non-simultaneous hard and soft flares,
we adopt the use of double filtering, i.e. discarding the time periods when either the hard-band or the soft-band light curves show excess by
more than 20\% over the nominal rate. If using only the outer part of the FOV (12--15 arcmin) the cluster contribution is small enough for the soft band filtering to be 
feasible.

The method for quiescent background estimation in our work is to use the filtered average blank sky spectrum in a given region, and to renormalize it 
to match the hard band count rate of a given observation. The blank sky data show that there is a poor correlation between the hard band (above 10 keV) and the 
useful 0.8--2.0 keV band, indicative of a remaining soft background component. Further, we found that when the hard band rate exceeds the average by more than 10\%, 
the 2.0--7.0 keV band excess remains at 10\%, indicative of an additional background component with harder spectrum than that of the quiescent particle background.
Thus, in order to minimize the uncertainties of our modeling, we adopt a method of not renormalizing the background below 2 keV, and for the higher energies, using a 
maximum of 1.1 for the renormalization factor.

Comparing the above models with the blank sky data, we determined the accuracy of the background modeling.
Using the double-filtering method, we obtain a 90\% CL uncertainty for the
model background rate of $\pm$ 5--20\%, for both MOS and
PN. Indicative of the higher variability of the particle-induced background at lower energies, 
the accuracy decreases from $\pm5$\% at $E=4-7$ keV to $\pm20$\% at $E=0.8-1$ keV. 
This compares to a $10-30$\% respective PN uncertainty when only the
hard-band light curve is used for filtering, and to a $15-45$\% PN uncertainty 
when applying the commonly used $2-3\sigma$ filtering method.
The improvement in the background modeling accuracy, when adding the soft-band filter,
results in only a 5--10\% reduction of the useful exposure time. 

Based on our analysis of a nearby bright cluster of galaxies A1795, the effect of the background uncertainties
on the temperature measurement, increase from a few \% at 3--4 arcmin to 20--25\% (i.e. $\sim$ 1 keV for A1795) at 10--15 arcmin. 
A conservative approximation of the commonly used 2--3 $\sigma$ clipping method using the hard band light curve in XMM-Newton works, 
yields that the resulting increased background uncertainties due to the allowed spectrally variable flare emission, 
render the temperatures essentially unconstrained at radii above above 10 arcmin in A1795.  
Thus, the background uncertainties do not prohibit the EPIC temperature profile analysis of low-brightness regions, like outer regions of galaxy clusters,
provided a conservative flare filtering such as the double filtering method with $\pm$20\% limits is used.

\acknowledgements

We thank Drs.\ J.\ Huovelin and H. Katayama and the referee, Dr. S. Molendi
for useful comments.  JN acknowledges support by NASA grants NAG5-9945 and
NAG5-13250; MM was supported by NASA contract NAS8-39073. Thanks to Dr
O. Vilhu for his support. Thanks to Meril\"a family for hospitality.

\clearpage

{\appendix

\section{Background subtraction methods}
\label{backmeth}

Here we summarize our method for spatially resolved spectroscopy of
extended sources in the 0.8--7 keV band.  It can be used for
observations with Thin and Medium optical filters in Full frame and
Extended full frame readout modes.

1) {\em Flare filtering.}\/ We use the full FOV, hard-band ($E>10$ keV
for PN, $E>9.5$ keV for MOS) cluster light curve (patterns 0--4 for PN, 0-12
for MOS) to determine a first-approximation quiescent level as a count
rate averaged in the range 0.54--0.80 cts s$^{-1}$ (PN) and 0.12--0.24 cts
s$^{-1}$ (MOS1 and MOS2).  All light curves should be binned into 1 ks bins to
ensure sufficient statistics. If there are no time bins where the count
rate is within the above ranges, the background differs significantly from
that in our sample, and thus our results cannot be applied. If any
inactive detector regions or the peculiar CCDs significantly reduce the size
of the useful detector area (from 615 arcmin$^{-2}$ for PN and 656
arcmin$^{-2}$ for MOS1 and MOS2 that we used in the blank sky sample), one needs to
reduce the above count rate limits accordingly. We then determine the
hard-band quiescent periods GTI$_h$ allowing the light curve to vary within
$\pm$20\% of the quiescent level.

Then we use the $r=12-15'$ (from the center of the FOV), $E=1-5$
keV light curve to determine the soft-band quiescent level. This value
contains source emission, and therefore varies depending on the source.
We use the Gaussian peak in the faint end of the count rate distribution
to determine the quiescent level.  If there are not enough time bins for a
proper distribution, we define the quiescent level so that the lower
20\% limit matches the minimum count rate. The soft-band quiescent
periods GTI$_s$ are determined by allowing the count rate to vary around the
above-determined value (that may include source emission) within $\pm$20\%
of the absolute values of the average blank-sky soft-band count rates of
$\pm$0.056 cts s$^{-1}$ (PN), or $\pm$0.028 cts s$^{-1}$ (MOS1 and MOS2). 
If these limits are smaller than the 2$\sigma$ statistical
uncertainties of the quiescent level (in the case of a very bright
source), one has to use only the hard band filter.  We then apply
both GTI filters simultaneously, e.g., using the following option in {\tt
evselect}: expression='GTI(GTI$_{h}$,TIME)\&\&GTI(GTI$_{s}$,TIME)'. 

A blank-sky dataset cleaned of flares in exactly such a manner is
available at \\
http://www.astro.helsinki.fi/$\sim$jnevalai/XMMbkg/.

2) {\em Background.}\/ We calculate the sky coordinates of the combined
blank-sky event file to match those of the cluster pointing, and
extract the blank-sky spectra in the same sky regions as used for the
data. The MOS1 and MOS2 spectra are co-added for the source as well as
the background.

3) {\em Background normalization.}\/ We compute a background
renormalization factor 
by obtaining the count rate in the hard band (10--14 keV
for PN, 9.5--12 keV for MOS), full FOV, flare-filtered spectrum of the
source in question, and dividing it by the corresponding value in the
co-added blank-sky spectrum.  This factor is applied to the background
model at $E>2$ keV. If it exceeds a factor of 1.1, we use 1.1 in this
band. For $E<2$ keV, we always use 1.0 (see \S\ref{method}).  For PN
Full frame data, we reduce the obtained renormalization factor by 6.3\%
to account for the background fraction subtracted as part of the OOT
correction. 
If the size of the accumulation region of the source and the
background region differs significantly, we adjust the renormalization factor
further.\footnote{Note that the $backscale$ tool in SAS distribution xmmsas\_20030110\_1802-5.4.1 yields correct values for the provided
co-added blank sky event file only when using the region expressions in detector coordinates (DETX,DETY).}

4) {\it OOT correction.}\/ For PN Full-frame data, we generate a simulated
OOT event file and extract the spectra with the same GTI and region
choices as for the source, and subtract the properly normalized OOT
spectra from the source spectra.

5) {\it Spectral fitting.}\/ In XSPEC, we input the source spectrum as
four separate datasets, and in each one, use only photons in one of the four
bands 0.8--1--2--4--7 keV (excluding the 1.45--1.55 keV band due to variable Al K lines). 
We set the BACKSCAL keyword in the background spectrum files to the
inverse of the renormalization factors calculated at step 3 (note that they are
different for $E<2$ keV and $E>2$ keV). Model parameters should then be
tied equal in all four sets to obtain a best-fit temperature and its
statistical uncertainty.

6a) {\it Systematic uncertainty, approximate estimate.}\/ Keeping the set-up from 5),
we input the original blank-sky spectrum in XSPEC as ``corfile'' for each of the four
datasets, and fit the data with all ``cornorm'' values set to 0.1 and then to
-0.1. The resulting difference in the best-fit temperature can be taken as a
90\% CL estimate of the systematic uncertainty. This method may overestimate the 
background uncertainties, especially for MOS data at large radii. Thus, if smaller
uncertainties are desired, one may use step 6b instead. If this accuracy is sufficient, instead of 4 energy
bands at step 5, one can use just two bands, 0.8--2--7 keV.

6b) {\it Systematic uncertainty, exact estimate.}\/ Instead of using the
same ``cornorm'' of $\pm0.1$ in all bands, we can use the uncertainty
correlation information to change the ``cornorm'' values in each of the
four bands individually, calculated from our 10 ks blank-sky pieces
(see the above website for this dataset). A distribution of the best-fit
temperatures obtained using each set of ``cornorms'' can then be used to
derive the 90\% CL interval for the systematic uncertainty.

\newpage

\section{Background components at $E<2$ keV}

Here we examine the EPIC background components at the lowest energies.
Below 2 keV, the sky background becomes important, compared to the particle
induced background.  The sky background consists mainly of Galactic
background and the extragalactic emission due to unresolved point sources.
The extragalactic component above E=0.8 keV is spatially uniform and
has a spectral shape of $\alpha_{ph}$ = 1.4. However, the Galactic
emission and absorption vary between the different pointings.

To estimate the contribution of different background components at
$E<2$ keV in our sample, we analyzed the co-added full FOV PN and MOS double
filtered quiescent spectra (from \S\ref{softfil}).  After subtracting
the co-added closed-cover spectrum (\S\ref{closed}), we modeled the
sky background spectrum, following Snowden et al.\ (1998), as a power-law
with the photon index of 1.4 (extragalactic background) and two thermal
components.  The best fit (Fig.\ \ref{f2.fig}) shows that at 1.0 keV
the extragalactic emission is comparable to that induced by the cosmic rays,
while the Galactic emission is negligible (note that all our blank-sky
fields are outside the bright Galactic features).  Towards lower energies
the Galactic emission becomes significant: at 0.8--1.0 keV, it comprises
15\% of the total background emission.  

We used the ROSAT All Sky Survey data (Snowden et al.\ 1997) to obtain
the 0.6--1.2 keV surface brightness for the different pointing
directions in our sample, obtaining a standard deviation and maximum
deviation from the mean of 10\% and $\pm$20\%, respectively. This level of
variation in the total sky background is substantial, but does not account
fully for the variation at these low energies that we observe.  Thus, in
addition to the variation in the Galactic emission, there should be
another variable emission component contributing significantly at energies
below 1 kev in our sample, probably residual background flares with a soft
spectrum.

\newpage

\section{Out-of-time correction to the blank sky data}
Since the PN Full frame source spectra must be corrected for the out-of-time (OOT) effect, for consistence this correction should also be applied to the 
blank sky data (the exact method). However, we assumed in this work that the net OOT effect on the background emission in a given
region can be approximated by the increase of the normalization by 6.3\% in the Full frame mode. This approximation was motivated by the
small fraction of the OOT events, and by the spatial and spectral uniformity of the background emission.
Here we examine in detail the accuracy of this approximation.

First, to obtain results using the exact method,  
we separated a Full frame mode subset from our
blank sky sample before co-adding the individual blank sky spectra. We produced
the OOT event files corresponding to these Full frame mode blank sky pointings, and co-added them into a single OOT event file. 
Using the same selection criteria as for the observed background spectra, and the co-added Full frame OOT event file, we extracted the 
OOT blank sky spectra.
After scaling them by 0.063, we subtracted them from the observed background spectra, producing
thus OOT corrected background spectra for the Full frame mode. 
We used the hard band of the full FOV spectrum
of the OOT corrected background and A1795 to obtain the renormalization
factor. After adjusting the background spectra with this factor, we subtracted them 
from the OOT corrected A1795 data and fitted the resulting spectra.

Then, we approximated the above
exact procedure by following the method used in this paper, i.e using the total co-added blank sky sample (including data obtained
in both Full frame and Extended full frame modes) without performing
any OOT correction. The subtraction of the hard-band-renormalized (see
above), not-OOT-corrected blank sky spectrum oversubtracts the
background, because both the simulated out-of-time emission of A1795 and
of the co-added blank sky spectra contain a out-of-time contribution due
to the background. Thus, before subtracting the not-OOT-corrected blank sky
spectra from OOT-corrected A1795 spectra, we reduced the background normalization by 6.3\%.
Fitting the data shows that the obtained temperatures agree with those derived
using the above exact method within 0.1 keV at all radii. Thus, it is justified to
reduce the complexity of the analysis with the approximation discussed here,
i.e. approximating the OOT effect on the background by reduction of the normalization of the not-OOT-corrected background 
by 6.3\% for any Full frame mode observation.}

\clearpage

\begin{figure}
\plotone{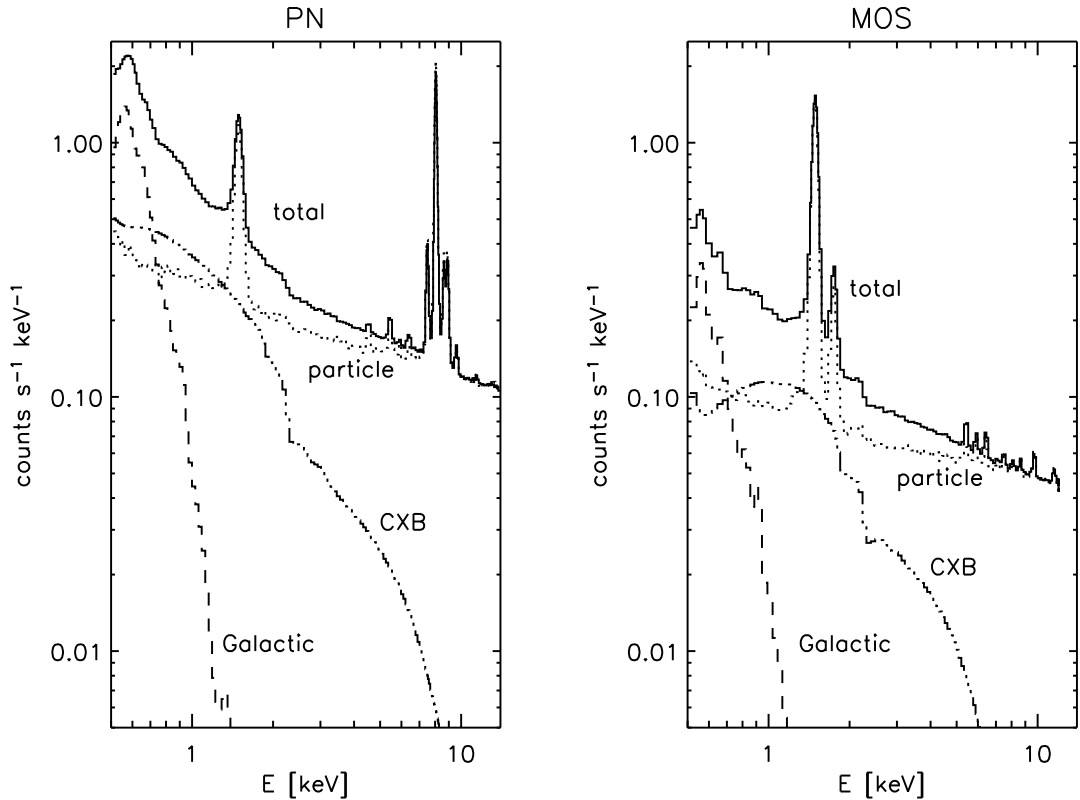}
\caption[]{The total quiescent PN and MOS background and convolved
  components as obtained from the full FOV spectra of the co-added
  closed-cover data sets (``particle'') and from the double-filtered (see
  \S\ref{softfil}) blank sky data sets (``CXB'' and ``Galactic'')}
\label{f2.fig}
\end{figure}

\end{document}